\documentclass[11pt]{article}

\usepackage[T1]{fontenc}
\usepackage[utf8]{inputenc}
\usepackage{amsmath,amssymb,amsthm,mathtools}
\IfFileExists{newtxtext.sty}{\usepackage{newtxtext,newtxmath}}{\usepackage{mathptmx}}
\usepackage[letterpaper,margin=0.92in]{geometry}
\usepackage{microtype}
\usepackage{booktabs,tabularx,array}
\usepackage{enumitem}
\usepackage{xcolor}
\usepackage{tikz}
\usetikzlibrary{quantikz2}
\let\ket\relax\let\bra\relax 
\usepackage{pgfplots}
\pgfplotsset{compat=1.18}
\usepackage{aliascnt}
\usepackage{hyperref}
\usepackage[nameinlink,capitalise,noabbrev]{cleveref}
\usepackage[numbers,sort&compress]{natbib}

\usepackage[nothm]{shared/cubicmoment}

\hypersetup{
  colorlinks=true,
  linkcolor=blue!55!black,
  citecolor=blue!55!black,
  urlcolor=blue!55!black,
  linktoc=all,
  pdftitle={Ancilla-free randomized Toffoli gates and the T-count of Boolean oracles},
  pdfauthor={Arul Rhik Mazumder},
}

\setlist[itemize]{leftmargin=1.5em,itemsep=2pt,topsep=3pt}
\setlist[enumerate]{leftmargin=1.7em,itemsep=2pt,topsep=3pt}
\allowdisplaybreaks
\makeatletter
\def\thm@space@setup{%
  \thm@preskip=4pt plus 1pt minus 2pt
  \thm@postskip=\thm@preskip
}
\makeatother

\newtheorem{theorem}{Theorem}

\newaliascnt{lemma}{theorem}
\newtheorem{lemma}[lemma]{Lemma}
\aliascntresetthe{lemma}
\crefname{lemma}{lemma}{lemmas}
\Crefname{lemma}{Lemma}{Lemmas}
\newaliascnt{proposition}{theorem}
\newtheorem{proposition}[proposition]{Proposition}
\aliascntresetthe{proposition}
\crefname{proposition}{proposition}{propositions}
\Crefname{proposition}{Proposition}{Propositions}
\newaliascnt{corollary}{theorem}
\newtheorem{corollary}[corollary]{Corollary}
\aliascntresetthe{corollary}
\crefname{corollary}{corollary}{corollaries}
\Crefname{corollary}{Corollary}{Corollaries}
\theoremstyle{definition}
\newaliascnt{definition}{theorem}
\newtheorem{definition}[definition]{Definition}
\aliascntresetthe{definition}
\crefname{definition}{definition}{definitions}
\Crefname{definition}{Definition}{Definitions}
\newaliascnt{example}{theorem}

\aliascntresetthe{example}
\crefname{example}{example}{examples}
\Crefname{example}{Example}{Examples}
\newaliascnt{construction}{theorem}
\newtheorem{construction}[construction]{Construction}
\aliascntresetthe{construction}
\crefname{construction}{construction}{constructions}
\Crefname{construction}{Construction}{Constructions}
\theoremstyle{remark}
\newaliascnt{remark}{theorem}
\newtheorem{remark}[remark]{Remark}
\aliascntresetthe{remark}
\crefname{remark}{remark}{remarks}
\Crefname{remark}{Remark}{Remarks}

\theoremstyle{plain}

\newcommand{\E}{\mathbb{E}}
\newcommand{\Prb}{\mathop{\mathrm{Pr}}}
\newcommand{\Tr}{\operatorname{Tr}}

\newcommand{\RPDT}{\mathrm{RPDT}^{\mathrm{na}}}

\newcommand{\Tmix}{\mathcal{T}^{\mathrm{mixed}}}

\newcommand{\Toff}{\mathrm{Toff}}

\newcommand{\OR}{\mathrm{OR}}
\newcommand{\AND}{\mathrm{AND}}
\newcommand{\eps}{\varepsilon}

\newcommand{\norm}[1]{\left\lVert #1\right\rVert}
\newcommand{\abs}[1]{\left\lvert #1\right\rvert}

\title{\bfseries Ancilla-free randomized Toffoli gates and the $T$-count of Boolean oracles}

\author{%
  Arul Rhik Mazumder\thanks{Imperial College London.
    \texttt{rhik.mazumder26@imperial.ac.uk}}
}
\date{}

\begin{document}
\maketitle

\begin{abstract}
In fault-tolerant Clifford+$T$ circuits, a Boolean oracle costs $T$ gates and extra qubits, and on four or more qubits every exact Toffoli gate needs an extra qubit.  Gosset, Kothari and Zhang showed that sampling a fresh random circuit for each call makes the $T$-count of the Toffoli gate independent of its size, but a direct implementation of their sampler uses about one extra qubit per bit of precision.  We show that computing the sampled parities in place, on the gate's own controls, removes every extra qubit once the gate is large enough: each random circuit then uses about $8$ $T$ gates per bit of precision and borrows only the gate's own idle inputs.  This is one $T$ gate more per circuit than the best measurement-free variant that keeps one clean ancilla.  The same holds for exclusive-ors of indicators of affine subspaces, such as equality tests and checks of linear equations.  For randomized circuits that do not measure, this is within a factor of eight of optimal up to an additive constant, even with extra qubits, and when each random circuit computes a Boolean function exactly, closing the factor asymptotically is equivalent to determining the leading constant in the $T$-count of exact multi-controlled Toffoli gates.  Finally, the oracle of the indicator of a random set of fixed density needs exponentially many $T$ gates at constant error, while that of an affine subspace of the same density needs a constant number, and extra qubits reduce the exponential cost only quadratically.
\end{abstract}

\tableofcontents

\section{Introduction}
\label{sec:intro}

A fault-tolerant Boolean oracle $U_f\ket{x}\ket{b}=\ket{x}\ket{b\oplus f(x)}$ has two costs: its $T$ gates, each of which consumes a distilled magic state~\cite{FowlerGidney,Litinski}, and its ancilla qubits beyond the inputs.  Such oracles are the inner loop of Grover search~\cite{Grover1996} and of block encodings for the quantum singular value transform~\cite{GilyenSuLowWiebe2019}.

The Toffoli gate $\Toff_n$ has $n-1$ controls and one target, and on a computational-basis input it flips the target exactly when the control string $x\in\F_2^{n-1}$ is all ones.  For this gate the two costs pull against each other.  Once $n\ge4$, no Clifford+$T$ circuit on its $n$ qubits implements it exactly \cite{GilesSelinger2013,Maslov2016}, so an exact circuit needs an extra qubit.  A clean ancilla, which starts and ends in $\ket0$, brings the cost to about $8n$ $T$ gates, with or without a mid-circuit measurement \cite{KhattarGidney2024,KoleAssaadSchnittkaDrechsler2026}.  A borrowed qubit, which may be in any state but must be restored, costs about $16n$ \cite{ZindorfBose2024,KhattarGidney2024}.  Neither count can drop below linear in $n$, since an exact circuit needs at least $n$ $T$ gates even with measurement and feedforward \cite[Proposition~4.1]{Beverland2020}.  Approximation alone does not help: for large $n$, a unitary circuit within diamond distance less than $1/2$ of $\Toff_n$ still needs $n-2$ $T$ gates even with clean ancillae \cite[Theorem~2]{gosset2025multiqubittoffoliexponentiallyfewer}, where diamond distance is the worst-case distance between quantum channels.

\begin{figure}[h]
\centering
\begin{quantikz}[row sep={0.9cm,between origins}, column sep=0.4cm]
\lstick{$n-1$ controls} & \qwbundle{} & \gate{X^{\otimes(n-1)}} & \gate[2]{P} & & \gate[2]{P^{\dagger}} & \gate{X^{\otimes(n-1)}} & \\
\lstick{$\ket{0^{\ell}}$} & \qwbundle{} & & & \octrl{1} & & & \rstick{$\ket{0^{\ell}}$} \\
\lstick{target} & & & & \targ{} & & &
\end{quantikz}
\caption{A direct implementation of one branch of GKZ's sampler.  A slash marks a bundle of qubits, and $X^{\otimes(n-1)}$ flips every control.  The CNOT circuit $P$ writes the $\ell$ sampled parities into $\ell$ clean ancillae, and the open circle is a control active on $0$.  The multi-controlled NOT is the only non-Clifford gate, and $P^{\dagger}$ returns the ancillae to $\ket0$.}
\label{fig:gkz-direct}
\end{figure}
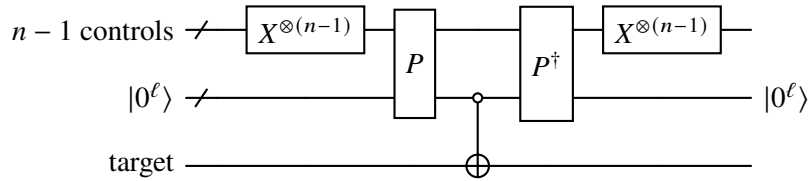

Gosset, Kothari and Zhang (GKZ) showed that randomness removes this dependence on $n$ \cite{gosset2025multiqubittoffoliexponentiallyfewer}.  If each call runs a freshly sampled Clifford+$T$ circuit and only the averaged channel has to be within diamond distance $\eps$ of $\Toff_n$, then $O(\log(1/\eps))$ $T$ gates suffice.  Their construction, which we call GKZ's sampler, first flips every control, so that the target must flip exactly when the control string is all zeros.  A parity of the controls is the sum modulo $2$ of a subset of them.  Each sampled circuit checks about $\log(1/\eps)$ parities with uniformly random subsets and flips the target only if all of them are $0$.  The all-zeros string passes every check, and any other string fails each check with probability $1/2$, so it passes them all with probability about $\eps$.  Once the parities are computed, the only non-Clifford gate is a multi-controlled NOT on their values.  A direct implementation computes them into fresh qubits (\Cref{fig:gkz-direct}) and so uses about $\log(1/\eps)$ ancillae, a number GKZ do not optimize.

We work in GKZ's mixed model (\Cref{def:models}).  A circuit implements a unitary $U$ if it applies $U$ up to a global phase.  A mixture is a probability distribution over unitary Clifford+$T$ circuits, called branches, which do not measure or adapt.  Its error is the diamond distance of the averaged channel from the target, its cost is the largest $T$-count of a branch, and $\Tmix_\eps(U)$ is the least cost at error $\eps$.  The oracle of $f:\F_2^n\to\{0,1\}$ acts on $n+1$ qubits, and $\Toff_n$ acts on $n$.  Our lower bounds allow any number of ancillae in any initial state, unless we say that the ancillae are clean.  Throughout, $\ell=\lceil\log(4/\eps)\rceil$ is the number of parities a branch checks, and logarithms are to base $2$.

Our results concern whether a mixture for $U_f$ still needs extra qubits (\Cref{sec:further}), how close it comes to the lower bounds (\Cref{sec:results}), and which property of $f$ determines its cost (\Cref{sec:random-intro}).

\begin{theorem}[the randomized Toffoli gate \informal{prop:toffoli-borrowed}]
\label{inf:toffoli}
If $n\ge\lceil3\ell/2\rceil$, then $\Toff_n$ has a mixture at diamond error $\eps$ whose branches use exactly $8\ell-8$ $T$ gates, no ancilla and no measurement.  If only $n\ge\ell+2$, the branches still use no ancilla and at most $16\ell-32$ $T$ gates.
\end{theorem}

The extra qubit is forced by a determinant.  Every Clifford+$T$ circuit on four or more qubits implements a unitary of determinant $1$, while an oracle that flips the target on $N$ control strings has determinant $(-1)^N$.  An exact Toffoli gate flips it on one string.  A branch that checks $\ell$ linearly independent parities also flips the target on every other string whose $\ell$ parities are all $0$.  These strings form a subspace of dimension $n-1-\ell$, so there are $2^{n-1-\ell}$ of them.  This number is even when $\ell\le n-2$, so the unitary such a branch must implement has determinant $1$ and the obstruction does not apply.  The parities depend only on the span of the sampled subsets, so a CNOT change of basis can overwrite at most $\ell$ of the controls with them, and the branch runs on the gate's own qubits.  We draw the subsets linearly independent, which keeps GKZ's error bound and makes the count above exact (\Cref{con:toffoli-branch}).  The checks then become a multi-controlled NOT on these $\ell$ controls, which can borrow the other $n-1-\ell$ controls as work qubits.

Against a direct implementation of GKZ's sampler, the branches save about $\ell$ ancillae.  Computing the parities in place still leaves one clean ancilla if the multi-controlled NOT is built on it, as in the cheapest exact constructions.  Once $n\ge\lceil3\ell/2\rceil$, the branches remove that qubit by borrowing idle controls instead, at the cost of one to five $T$ gates (\Cref{tab:toffoli-resources}), which is worthwhile when a computation is limited by its number of logical qubits.  With $\ell-2$ clean ancillae and measurement, the sampler can instead use a ladder of temporary logical-AND gates \cite{Jones2013,Gidney2018} and needs $4\ell-1$ $T$ gates \cite{KoleSchnittkaDrechsler2026}, about half of $8\ell-8$.  We know of no other Clifford+$T$ implementation of $\Toff_n$ that uses no qubit beyond its own $n$ and a number of $T$ gates independent of $n$.  Once $n\ge\lceil3\ell/2\rceil$, the branches also use fewer $T$ gates than every exact construction without a clean ancilla, and they need no measurement.

\subsection{Structured oracles}
\label{sec:further}

A branch of GKZ's sampler flips the target exactly on the solutions of $\ell$ linear equations over $\F_2$, so it is itself a test of linear equations.  Many useful oracles have the same form.  The inputs that satisfy $r$ independent equations form an affine subspace of dimension $n-r$, that is, of codimension $r$.  CNOT gates can write the value of each equation onto one of the inputs, so the oracle of such a test is $\Toff_{r+1}$ up to Clifford gates, and our branches extend to it.  They also extend to an exclusive-or of $J$ such oracles, built term by term at error $\eps/J$ each.  We write $\mathbf1_A$ for the indicator function of a set $A$.

\begin{theorem}[affine decompositions \informal{prop:affine-xor}]
\label{inf:affine}
Let $f=\bigoplus_{j=1}^J\mathbf1_{A_j}$ with $A_j$ of codimension $r_j$, and put $\ell_J=\lceil\log(4J/\eps)\rceil$.  If $n\ge\lceil3\min\{r_j,\ell_J\}/2\rceil-1$ for every $j$, then $U_f$ has a mixture at diamond error $\eps$ with no ancilla, no measurement and at most $8J\ell_J$ $T$ gates per branch, independently of $n$.
\end{theorem}

With $J=1$, \Cref{inf:affine} covers the reflection about $\ket{0^a}$, which is $\Toff_a$ up to Clifford gates and is used in amplitude amplification~\cite{Grover1996} and in the quantum singular value transform~\cite{GilyenSuLowWiebe2019}.  It also covers equality tests between two registers or against a fixed string.  Membership in the solution set of a linear system, or in a linear code, is another example.  For these oracles GKZ's construction already gives $O(\log(1/\eps))$ $T$ gates \cite{gosset2025multiqubittoffoliexponentiallyfewer}, and \Cref{inf:affine} removes its ancillae.

Larger $J$ covers functions that split into disjoint pieces.  The inputs that reach a leaf of a decision tree satisfy the conditions along its path, so they form an affine subspace.  Different leaves share no input, so the function is an exclusive-or of one indicator per accepting leaf.  The same holds for parity decision trees, which query parities of the input instead of single bits, and for disjunctive normal forms with disjoint terms.  For these functions \Cref{inf:affine} matches, up to constants, the bound that GKZ obtain with ancillae \cite[Theorem~5]{gosset2025multiqubittoffoliexponentiallyfewer}.

These savings shrink when the oracle is called many times.  If a computation makes $N$ calls, each call needs error about $\eps/N$, because diamond errors add up over a sequence of calls.  So $\ell$ grows like $\log(N/\eps)$, and the branches pay off only on oracles with somewhat more than $\log(N/\eps)$ qubits (\Cref{fig:crossover}).

\subsection{Lower bounds and exact gates}
\label{sec:results}

GKZ proved that mixtures for $\Toff_n$ need $\Omega(\min\{n,\log(1/\eps)\})$ $T$ gates, even with measurement and adaptivity \cite{gosset2025multiqubittoffoliexponentiallyfewer}.  They also showed that a mixture with $t$ $T$ gates can be simulated by a random choice of parity decision trees, each querying about $t$ parities of the input \cite[Theorem~23]{gosset2025multiqubittoffoliexponentiallyfewer}.  With its controls flipped, $\Toff_n$ computes the negation of the $\OR$ function, and a standard calculation gives the exact number of parity queries a random decision tree needs for $\OR$ (\Cref{prop:or-rpdt}).  The following explicit bound for unitary mixtures is a corollary of their reduction, with leading constant $1$, where the argument of their Theorem~3 gives about $1/4$ but also covers measurement and adaptivity.

\begin{theorem}[lower bound for the Toffoli gate \informal{prop:toffoli-lower}]
\label{inf:lower}
For $0<\eps\le1/2$, every mixture of unitary Clifford+$T$ circuits that approximates $\Toff_n$ to diamond error $\eps$ has a branch with at least $\min\{n-1,\log((1-\eps)/\eps)\}-2$ $T$ gates.
\end{theorem}

By \Cref{inf:lower}, the branches of \Cref{inf:toffoli} are within a factor of eight of optimal up to an additive constant (\Cref{cor:factor-eight}).  The same holds for GKZ's sampler built on the best known clean-ancilla multi-controlled NOTs.  In both constructions every branch is the exact oracle of some Boolean function, though not the one $\Toff_n$ computes.  For mixtures of this kind, \Cref{inf:oracle} shows that closing the factor of eight is equivalent to finding exact multi-controlled NOTs with fewer $T$ gates.  Let $t(k)$ be the least $T$-count of an exact $k$-controlled NOT with clean ancillae and no measurement.  The stabilizer-nullity bound of Beverland et al.\ and Maslov's construction give $k+1\le t(k)\le8k-9$ for $k\ge3$ \cite[Proposition~4.1]{Beverland2020}, \cite[Proposition~4]{Maslov2016}, and we know of no better bounds.

\begin{theorem}[oracle branches and exact multi-controlled NOTs \informal{thm:oracle-equivalence}]
\label{inf:oracle}
Let $0<\eps\le1/2$ and $n-1\ge\ell$.  Among mixtures of circuits that each implement some Boolean oracle exactly, with any clean ancillae, the least branch $T$-count for $\Toff_n$ at diamond error $\eps$ is between $t(K)-8$ and $t(K)+24$, where $K=\lfloor\log(1/\eps)\rfloor$.
\end{theorem}

GKZ's sampler gives the upper bound \cite[Theorem~1]{gosset2025multiqubittoffoliexponentiallyfewer}, and the lower bound is new.  So without a better exact multi-controlled NOT, closing the factor of eight needs branches that are not the exact oracle of any Boolean function, for example circuits that add input-dependent phases (\Cref{thm:oracle-equivalence}).

\subsection{Distance from affine and random sets}
\label{sec:random-intro}

An affine function is a parity of some of the inputs or its negation, and its oracle needs no $T$ gates, since CNOT and $X$ gates implement it \cite{LiGanWu2025stabqram}.  The function computed by $\Toff_n$ differs from the affine function $0$ on a single input, and by the intuition behind \Cref{inf:lower} that input is what costs $T$ gates.  This suggests measuring a general $f$ by its distance from affine.  Let $\delta$ be the smallest fraction of inputs on which a non-affine $f$ differs from an affine function, and call those inputs exceptional.  Through GKZ's reduction to decision trees, $f$ needs at least $\log(1/\max\{\delta,2\eps\})-3$ $T$ gates (\Cref{thm:distance}).  When the exceptional inputs form an affine subspace, this is tight up to a factor of eight and an additive constant (\Cref{cor:distance-two-sided}).  At constant error the bound is only a constant, and the cost then depends on how the exceptional inputs are arranged.  \Cref{inf:unstructured} compares the two extremes, a subspace and a set scattered at random.

\begin{theorem}[distance does not determine the $T$-count \informal{cor:distance-not-enough}]
\label{inf:unstructured}
Let $6\le k\le n-3\log n-7$.  Let $A$ be an affine subspace of codimension $k$, and let $R$ be a random set that contains each input independently with probability $2^{-k}$.  With probability at least $1-2^{2-n}$ over $R$, the following hold at $\eps=1/4$.
\begin{enumerate}[nosep,label=(\arabic*)]
\item Both $\mathbf1_A$ and $\mathbf1_R$ are at distance between $\frac45 2^{-k}$ and $\frac65 2^{-k}$ from affine.
\item The oracle of $\mathbf1_A$ costs at most $24$ $T$ gates.
\item The oracle of $\mathbf1_R$ costs $2^{(n-k)/2\pm O(\log n)}$ $T$ gates with clean ancillae and $2^{n-k\pm O(\log n)}$ without ancillae.
\end{enumerate}
\end{theorem}

Both functions are $0$ on almost every input, so their exceptional inputs are $A$ and $R$ themselves, each about a $2^{-k}$ fraction of the inputs.  They differ only in structure.  Membership in $A$ is decided by $k$ parities, and at $\eps=1/4$ a branch of \Cref{inf:affine} checks just four random combinations of them.  A random set has no such description: it meets every large subspace in about its expected fraction, so no few parities single it out.  Counting makes this precise, since mixtures with few $T$ gates can approximate far fewer functions than there are sets $R$.

The upper bounds hash the input to about $n-k$ bits and implement the hashed set exactly (\Cref{prop:random-upper}), and the lower bounds count the functions that cheap mixtures can approximate (\Cref{thm:random-exponential}).  With clean ancillae the count adapts that of Gosset, Kothari and Wu \cite[Lemma~4.8]{GKW24}, and without ancillae a finer count gives the larger bound.  That count extends the exact no-ancilla counting bound of Low, Kliuchnikov and Schaeffer \cite[Section~5]{LKS24} to mixtures at constant error.

Some sets of this density were already known to be expensive.  Gosset, Kothari and Wu's counting gives a function on $n-k$ bits that needs $\Omega(2^{(n-k)/2})$ $T$ gates even with measurement, and requiring the other $k$ inputs to be $0$ gives one of density about $2^{-k}$.  Earlier counting arguments for sparse sets apply only to exact circuits.  \Cref{inf:unstructured} shows that a typical set of this density is expensive even at diamond error $1/4$ and that matching upper bounds exist, so distance from affine does not determine the cost.  For an oracle with no structure, randomness cannot remove the exponential cost, and clean ancillae reduce it only from about $2^{n-k}$ to about $2^{(n-k)/2}$.  An oracle whose exceptional inputs form an affine subspace stays cheap and needs no ancillae, by \Cref{inf:affine}.

\Cref{sec:overview} goes through one branch of the Toffoli construction and sketches the proofs, \Cref{sec:related} discusses related work, and \Cref{sec:discussion} the open problems.  Appendix~\ref{sec:prelim} fixes notation and the model and reviews the multi-controlled NOTs we build on, Appendix~\ref{sec:constructions} proves the constructions, and Appendix~\ref{sec:lower} proves the lower bounds.  Python scripts that check the constructions and the numerical claims are available at \url{https://github.com/arulrhikm/randomized-oracle-checks}.
\section{Technical overview}
\label{sec:overview}

\subsection{Warm-up: one branch}
\label{sec:warmup}

We first make the branch from \Cref{sec:intro} precise.  Write $y=x\oplus\mathbf1$ for the control string with every bit flipped, so that $\Toff_n$ flips the target exactly when $y=0$.  A branch picks subsets $S_1,\ldots,S_\ell$ of the controls whose indicator vectors are linearly independent over $\F_2$, uniformly at random among such choices.  It flips the target when $y$ has even parity on every $S_i$.  A nonzero $y$ passes all $\ell$ tests with probability at most $2^{-\ell}$ (\Cref{con:toffoli-branch}), as it does for GKZ's uniformly random subsets.  A mixture of oracles that errs on each input with probability at most $p$ is within diamond distance $4p$ of the target (\Cref{lem:pointwise-miss}), so $\ell=\lceil\log(4/\eps)\rceil$ gives error at most $\eps$.  When $\ell\le n-2$, the determinant obstruction of \Cref{sec:intro} does not apply (\Cref{prop:determinant}).

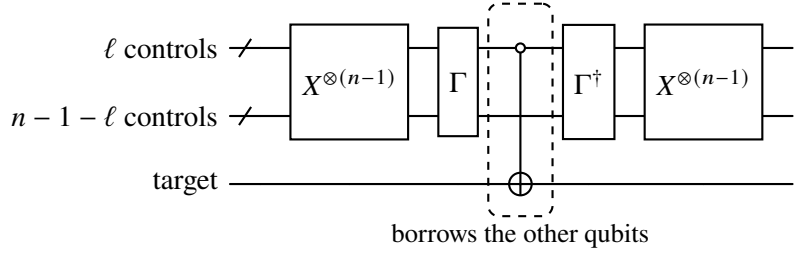
\begin{figure}[!htb]
\centering
\begin{quantikz}[row sep={0.9cm,between origins}, column sep=0.4cm]
\lstick{$\ell$ controls} & \qwbundle{} & \gate[2]{X^{\otimes(n-1)}} & \gate[2]{\Gamma} & \octrl{2}\gategroup[3,steps=1,style={dashed,rounded corners,inner xsep=4pt},label style={label position=below,anchor=north,yshift=-0.15cm}]{\small borrows the other qubits} & \gate[2]{\Gamma^{\dagger}} & \gate[2]{X^{\otimes(n-1)}} & \\
\lstick{$n-1-\ell$ controls} & \qwbundle{} & & & & & & \\
\lstick{target} & & & & \targ{} & & &
\end{quantikz}
\caption{One branch of \Cref{inf:toffoli}, drawn as in \Cref{fig:gkz-direct} but with no ancilla.  The CNOT circuit $\Gamma$ moves the $\ell$ sampled parities onto the first $\ell$ controls.  The multi-controlled NOT uses the other $n-1-\ell$ qubits as borrowed work qubits and returns them unchanged, and it is the only non-Clifford gate in the branch.}
\label{fig:toffoli}
\end{figure}

Completing the subsets to a basis gives an invertible CNOT circuit $\Gamma$ whose first $\ell$ outputs are the parities of $y$ on $S_1,\ldots,S_\ell$.  Applying $\Gamma$, then an $\ell$-controlled NOT with controls active on $0$, then $\Gamma^{\dagger}$ implements the branch with no extra qubit (\Cref{fig:toffoli}).  The multi-controlled NOT does not depend on the other $n-1-\ell$ qubits.  A circuit for it can therefore use them as borrowed work qubits, which may be in any state and are returned unchanged, as in the constructions of Barenco et al.\ \cite[Lemma~7.2]{barenco1995}.  Maslov's construction uses $8\ell-8$ $T$ gates once $\lceil(\ell-2)/2\rceil$ such qubits are available, that is, once $n\ge\lceil3\ell/2\rceil$ \cite[Proposition~5]{Maslov2016}.  It is built from relative-phase Toffoli gates, whose phases cancel in pairs (\Cref{sec:mcx}).

\begin{figure}[!tb]
\centering
\begin{tikzpicture}
\begin{axis}[
  width=0.78\textwidth, height=4.6cm,
  xlabel={$n$}, ylabel={$T$-count of $\Toff_n$},
  xmin=5, xmax=100, ymin=0, ymax=420,
  legend columns=3, legend cell align=left,
  legend style={font=\scriptsize, at={(0.5,-0.34)}, anchor=north, /tikz/every even column/.append style={column sep=0.8em}},
  tick label style={font=\scriptsize}, label style={font=\small}]
\input{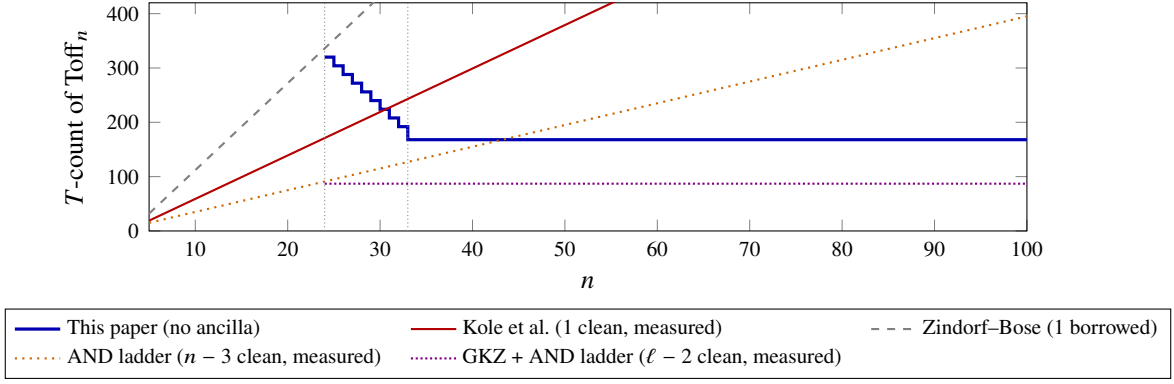}
\draw[densely dotted, gray] (axis cs:24,0) -- (axis cs:24,420);
\draw[densely dotted, gray] (axis cs:33,0) -- (axis cs:33,420);
\legend{{This paper (no ancilla)},{Kole et al.\ ($1$ clean, measured)},{Zindorf--Bose ($1$ borrowed)},{AND ladder ($n-3$ clean, measured)},{GKZ + AND ladder ($\ell-2$ clean, measured)}}
\end{axis}
\end{tikzpicture}
\caption{$T$-count against $n$ at $\eps=10^{-6}$, where $\ell=22$, for the constructions of \Cref{tab:toffoli-resources}.  Dotted lines mark $n=\ell+2=24$ and $n=\lceil3\ell/2\rceil=33$.  Below $n=24$ the branches need a clean ancilla and are not drawn.  Between the lines they cost $16\ell-16d-16$ with $d=n-1-\ell$ idle controls, and from $n=33$ on they cost $168$.  The exact circuits with at most one ancilla leave the frame at $n=30$ and $n=56$, and GKZ + Khattar--Gidney, which is not drawn, lies one $T$ gate below the branches.}
\label{fig:crossover}
\end{figure}

\Cref{tab:toffoli-resources} and \Cref{fig:crossover} compare the $T$-counts of the branches of \Cref{inf:toffoli} with those of GKZ's sampler built on three other multi-controlled NOTs and of exact constructions.  Measurement and reset are counted as free.  Every other construction in the table uses an extra qubit.  Those without a clean ancilla need at least $384$ $T$ gates, against $88$ to $352$ for the branches.  For $n$ between $\ell+2$ and $\lceil3\ell/2\rceil-1$, the branches use Zindorf and Bose's construction instead, which borrows the $d=n-1-\ell$ idle controls \cite{ZindorfBose2024}.  Even there they can beat the exact circuits with one clean ancilla, for example with $352$ against $379$ $T$ gates at $n=50$ and $\eps=10^{-10}$.  From $n=\ell+2$ the branches need no ancilla and use fewer $T$ gates than exact circuits with one borrowed qubit.  They beat exact circuits with one clean ancilla once $n>4\ell/3+1$, and those with $n-3$ clean ancillae once $n\ge2\ell$.

\begin{table}[!t]
\centering\small
\renewcommand{\arraystretch}{1.12}
\resizebox{\textwidth}{!}{%
\begin{tabular}{@{}lllccrrr@{}}
\toprule
 & & & & & \multicolumn{3}{c}{$T$-count at $n=50$} \\
\cmidrule(l){6-8}
Construction & Source & Ancillae & Meas. & $T$-count & $\eps=10^{-3}$ & $\eps=10^{-6}$ & $\eps=10^{-10}$ \\
 & & & & & $(\ell=12)$ & $(\ell=22)$ & $(\ell=36)$ \\
\midrule
\multicolumn{8}{@{}l}{\emph{Randomized, diamond error $\eps$}} \\
This paper & \Cref{inf:toffoli} & none (idle controls) & no & $8\ell-8$ & $\mathbf{88}$ & $\mathbf{168}$ & $\mathbf{352}^{\dagger}$ \\
GKZ + Khattar--Gidney & \cite{gosset2025multiqubittoffoliexponentiallyfewer,KhattarGidney2024} & $1$ clean & no & $8\ell-9$ & $87$ & $167$ & $279$ \\
GKZ + Kole et al. & \cite{gosset2025multiqubittoffoliexponentiallyfewer,KoleAssaadSchnittkaDrechsler2026} & $1$ clean & yes & $8\ell-13$ & $83$ & $163$ & $275$ \\
GKZ + AND ladder & \cite{gosset2025multiqubittoffoliexponentiallyfewer,Jones2013,Gidney2018} & $\ell-2$ clean & yes & $4\ell-1$ & $47$ & $87$ & $143$ \\
\midrule
\multicolumn{8}{@{}l}{\emph{Exact}} \\
Zindorf--Bose & \cite[Theorem~2]{ZindorfBose2024}, \cite[Table~1]{KhattarGidney2024} & $1$ borrowed & no & $16m-32$ & \multicolumn{3}{c}{$752$} \\
Maslov & \cite[Proposition~5]{Maslov2016} & $\lceil(m-2)/2\rceil$ borrowed & no & $8m-8$ & \multicolumn{3}{c}{$384$} \\
Khattar--Gidney & \cite{KhattarGidney2024}, \cite[Eq.~(12)]{KoleAssaadSchnittkaDrechsler2026} & $1$ clean & no & $8m-9$ & \multicolumn{3}{c}{$383$} \\
Kole et al. & \cite[Eq.~(8)]{KoleAssaadSchnittkaDrechsler2026} & $1$ clean & yes & $8m-13$ & \multicolumn{3}{c}{$379$} \\
AND ladder & \cite{Jones2013,Gidney2018}, \cite[Eqs.~(3)--(4)]{KoleSchnittkaDrechsler2026} & $m-2$ clean & yes & $4m-1$ & \multicolumn{3}{c}{$195$} \\
\bottomrule
\end{tabular}}
\caption{$T$-counts for $\Toff_n$ with $m=n-1$ controls.  The branches are in bold.  GKZ + X is GKZ's sampler with its parities computed in place as in \Cref{con:toffoli-branch} and its multi-controlled NOT built as in X.  Computing them into fresh qubits instead, as in GKZ's direct circuit, adds $\ell$ ancillae.  The Khattar--Gidney rows use the count without measurement of \cite[Eq.~(12)]{KoleAssaadSchnittkaDrechsler2026}; with the measurement-based uncomputation of \cite{KhattarGidney2024} the count is $8m-12$.  At the entry marked $\dagger$, $n<\lceil3\ell/2\rceil$ and the count is $16\ell-16d-16$ with $d=n-1-\ell$ \cite[Theorems~2 and~10]{ZindorfBose2024}.}
\label{tab:toffoli-resources}
\end{table}

\subsection{Structured oracles}

\Cref{inf:affine} applies the branch of \Cref{sec:warmup} to each term of an affine decomposition, so the only new questions are which terms to randomize and how their errors add.  A term of codimension at most $\ell_J$ is implemented exactly, as in the synthesis of exclusive-or sums of affine products \cite{SchaefferPerkowski2013}.  A term of larger codimension is replaced by a branch that tests $\ell_J$ random combinations of its equations.  Such a term misses an input with probability at most $2^{-\ell_J}\le\eps/(4J)$.  The misses of the $J$ terms add, so each input is missed with probability at most $\eps/4$, and \Cref{lem:pointwise-miss} turns this into diamond error at most $\eps$.  Every Boolean function has such a decomposition, for instance with one term per accepting input, but $J$ can then be exponential in $n$.  A parity decision tree of depth $D$ with $J$ accepting leaves gives a decomposition that costs at most $8J\min\{D,\ell_J\}$ $T$ gates when $n\ge\lceil3\min\{D,\ell_J\}/2\rceil-1$ (\Cref{sec:general-decompositions}).

For example, the membership oracle of Grover search with $J$ marked strings is an exclusive-or of $J$ indicators of single points, so when $n\ge\lceil3\ell_J/2\rceil-1$ it uses $J(8\ell_J-8)$ $T$ gates and no ancilla.  Grover search makes about $2^{n/2}$ calls, so the error per call must be about $\eps_{\mathrm{tot}}2^{-n/2}$.  At $n=100$, $J=1$ and $\eps_{\mathrm{tot}}=10^{-2}$, this gives $\ell_1=59$, and each call costs $464$ $T$ gates on the oracle's own $101$ qubits.  The oracle is $\Toff_{101}$ up to Clifford gates, so \Cref{tab:toffoli-resources} applies with $m=100$.  An exact circuit needs an extra qubit.  Without measurement it uses $1568$ $T$ gates with a borrowed qubit \cite{ZindorfBose2024,KhattarGidney2024} or $791$ with a clean ancilla \cite{KhattarGidney2024}, \cite[Eq.~(12)]{KoleAssaadSchnittkaDrechsler2026}, which is $3.4$ and $1.7$ times as many.  Only constructions with clean ancillae do better: GKZ's sampler uses $463$ with one, and the AND ladder uses $399$ with $98$ and measurement.  So even at the error that $2^{50}$ calls demand, the branches free a qubit for one $T$ gate per call.

\subsection{Lower bounds for the Toffoli gate}

The proof of \Cref{inf:lower} goes through parity decision trees.  GKZ show that a mixture whose branches use $t$ $T$ gates can be simulated by a random choice of decision trees, each asking $t+1$ parities of the input and then outputting any function of the answers \cite[Theorem~23]{gosset2025multiqubittoffoliexponentiallyfewer}.  It therefore suffices to show that random trees with few queries cannot compute $\Toff_n$, which in the coordinates $y$ of \Cref{sec:warmup} is the negation of $\OR$ on $m=n-1$ bits.  Fix one tree that asks $q$ parities.  All $y$ in the kernel of these parities give the same answers, so the tree treats $y=0$ like the $2^{m-q}-1$ nonzero points of this kernel and errs either at $0$ or on all of them.  Put half the weight on $y=0$ and spread the other half over the nonzero inputs.  Every such tree then errs with probability about $2^{-q}/2$, and by Yao's principle so does every random choice of trees.  Error $\eps$ thus needs $q\approx\log(1/\eps)$, as in the standard argument for one-sided Equality \cite{KMSY18,MontanaroOsborne2009}.  The bound is tight: a mixture of GKZ's test with a tree of constant output meets it exactly (\Cref{prop:or-rpdt}), and with the reduction this gives \Cref{inf:lower}.

\Cref{inf:oracle} needs a different argument.  Decision-tree bounds stay below $\lceil\log(1/\eps)\rceil$, so they cannot see the constant in front of $\log(1/\eps)$.  The proof instead works with a single branch.  For mixtures of oracles, diamond error $\eps$ bounds the error probability at every input by $\eps$ (\Cref{lem:oracle-pointwise}), so some branch is correct at $y=0$ and wrong on few other inputs.  A random subspace of dimension about $\log(1/\eps)$ that contains $y=0$ then avoids all of those inputs with positive probability, and we restrict the branch to it.  What remains is an exact multi-controlled NOT on about $\log(1/\eps)$ controls.

\subsection{Distance from affine and random sets}

\Cref{thm:distance} extends the argument for \Cref{inf:lower} from $\Toff_n$ to any $f$.  Let $a$ be the affine function closest to $f$, so that $f$ and $a$ differ on a fraction $\delta$ of the inputs.  Since the oracle of $a$ needs no $T$ gates, implementing $U_f$ is as hard as implementing the oracle of $f\oplus a$, which is $1$ exactly on the exceptional inputs.  Put half the weight on the exceptional inputs and spread the other half over the rest.  A tree that asks $q$ parities gives the same answers on each coset of their kernel, which holds a $2^{-q}$ fraction of the inputs.  It therefore errs with probability at least $(2^{-q}-\delta)/2$, and by Yao's principle so does every random choice of trees.  Error $\eps$ thus needs about $\log(1/\max\{\delta,2\eps\})$ queries.

When the exceptional inputs form an affine subspace $A$ of codimension $r$, so that $f=a\oplus\mathbf1_A$ and $\delta=2^{-r}$, \Cref{inf:affine} builds the oracle from a single term with at most about $8\min\{r,\ell\}$ $T$ gates.  Against the lower bound above, this gives the factor of eight in \Cref{cor:distance-two-sided}.  An exact implementation needs at least $r+1$ $T$ gates, so randomization helps only when $\ell<r$, that is, roughly when $\delta\lesssim\eps/4$.

For \Cref{inf:unstructured}, a Chernoff bound puts $|R|$ within a factor of $6/5$ of $2^{n-k}$.  Every affine function other than $0$ differs from $\mathbf1_R$ on about half the inputs, so the distance of $\mathbf1_R$ from affine is about $2^{-k}$.  Decision-tree bounds stay below $n-k+3$ here, so the exponential bounds use counting instead.  Averaging $O(n)$ sampled branches and rounding recovers $\mathbf1_R$ exactly, so a mixture with few $T$ gates would give a short description of $R$.  Circuits with $t$ $T$ gates have only $2^{O(t^2+nt)}$ possible acceptance-probability functions, or $2^{O(nt)}$ without ancillae (\Cref{lem:acceptance-count}).  There are $2^{\Omega(2^{n-k})}$ possible sets $R$, so $t$ must be about $2^{(n-k)/2}$ with clean ancillae and about $2^{n-k}$ without.

The upper bounds hash the input to about $n-k$ bits with a random CNOT circuit and implement the oracle of the hashed set exactly.  An input outside $R$ lands in the hashed set only with small probability, and this is the error of the mixture.  With clean ancillae the hashed oracle follows the randomized construction of Li et al.\ for sparse lookup tables \cite[Theorem~3.4]{LiOuWangYaoYuanZhang2026}.  Without ancillae it is built point by point from Zindorf and Bose's multi-controlled NOT \cite{ZindorfBose2024}, which borrows one of the unused inputs.
\section{Related work}
\label{sec:related}

Campbell and Hastings showed that a random mixture of circuits can approximate a single-qubit rotation more cheaply than any one circuit, because the errors of the branches average out \cite{Campbell2017mixing,Hastings2017}.  Gosset, Kothari and Zhang carried the idea to multi-qubit Boolean oracles \cite{gosset2025multiqubittoffoliexponentiallyfewer}.  They gave $O(\log(1/\eps))$ $T$ gates for $\Toff_n$ and $O(\norm{\widehat f}_1^2\log(1/\eps))$ for an arbitrary $f$, where $\norm{\widehat f}_1$ is the sum of the absolute Fourier coefficients of $f$.  \Cref{inf:toffoli} computes their parities in place and borrows the idle controls, so each branch runs without ancillae, and \Cref{inf:affine} does the same for affine decompositions.

GKZ also proved a matching $\Omega(\min\{n,\log(1/\eps)\})$ lower bound by generalizing stabilizer nullity, and bounded $T$-counts below by randomized parity decision trees.  For mixtures of oracle branches, \Cref{inf:oracle} shows that their construction, built on an optimal exact multi-controlled NOT, is optimal up to an additive constant.  Girish, May, Parham and Yuen proved $T$-count lower bounds for mixtures through communication complexity, including with advice states \cite{GirishMayParhamYuen2025}.  Randomized non-adaptive parity decision trees are randomized linear sketches over $\F_2$, studied by Kannan et al.\ and by Montanaro and Osborne \cite{KMSY18,MontanaroOsborne2009}, and \Cref{prop:or-rpdt,prop:distance-rpdt} use the standard averaging arguments from that setting.

Counting arguments give lower bounds for unstructured oracles.  Low, Kliuchnikov and Schaeffer implemented the oracle of an arbitrary $n$-bit function with $O(\sqrt{2^n})$ $T$ gates using ancillae, and proved a counting lower bound when the number of ancillae is restricted \cite{LKS24}.  Gosset, Kothari and Wu settled the $T$-count of state preparation and diagonal unitaries by counting canonical forms whose size does not depend on the ancillae \cite{GKW24}.  As GKZ note, this gives a Boolean oracle that needs $\Omega(2^{n/2})$ $T$ gates at constant error, even with measurement.

Sparse oracles have been studied separately.  Li et al.\ determined the $T$-count of sparse classical data \cite{LiOuWangYaoYuanZhang2026}, Luo and Li synthesized sparse Boolean functions with $\widetilde O(n^{3/4}\sqrt s)$ Toffoli gates for a large support of size $s$ \cite{LuoLi2026sparse}, and Nie and Zi counted the total gates of sparse Boolean oracles \cite{NieZi2026boolean}.  For one-bit oracles of sparse sets, the counting lower bounds of Li et al.\ and of Nie and Zi need exact implementations.  Average-case error cannot capture these oracles, because such an oracle agrees with the identity on all but a small fraction of inputs.  The approximate bounds of Li et al.\ assume at least two output bits \cite[Theorem~5.1]{LiOuWangYaoYuanZhang2026}, although they allow measurement and feedforward.  \Cref{thm:random-exponential} instead holds at constant diamond error for mixtures of unitary circuits, with clean ancillae or without, and for a random set with high probability.

Exact constructions differ in the number and kind of extra qubits they use.  For exact Clifford+$T$ circuits one ancilla always suffices \cite{GilesSelinger2013}, as it does for reversible circuits of NOT, CNOT and Toffoli gates \cite{Shende2003}.  The determinant decides which oracles need that ancilla, namely those of full degree over $\F_2$ (\Cref{rem:degree-criterion}), and so rules out ancilla-free exact Toffoli gates on four or more qubits \cite{Maslov2016}.  Avoiding the ancilla exactly would also need gates of exponential precision \cite{NieZiSun2024}.  Randomized branches escape both obstructions, because no branch has to implement $\Toff_n$ exactly.  The exact constructions of \Cref{tab:toffoli-resources} fall into three groups.  Some borrow idle qubits \cite{barenco1995,Maslov2016,ZindorfBose2024}.  Others use one clean ancilla through conditionally clean ancillae \cite{NieZiSun2024,KhattarGidney2024,KoleAssaadSchnittkaDrechsler2026}.  The AND ladder uses many clean ancillae as temporary logical-AND gates \cite{Jones2013,Gidney2018,KoleSchnittkaDrechsler2026}.

Reversible-logic synthesis builds the oracle of a Boolean function from an exclusive-or of products of literals, one multi-controlled NOT per term.  Fazel, Thornton and Rice gave an early cascade construction \cite{FazelThorntonRice2007}, and Meuli et al.\ compiled low-$T$ oracles from such decompositions and from the multiplicative complexity of $f$ \cite{MeuliSoeken2019}.  Schaeffer and Perkowski allowed each factor to be a parity, so that each term is the indicator of an affine subspace \cite{SchaefferPerkowski2013}.  \Cref{inf:affine} implements those terms without ancillae and randomizes the terms of large codimension, as GKZ's sampler does for a single term.
\section{Discussion and open problems}
\label{sec:discussion}

We do not know the constant in the $T$-count of randomized Toffoli gates.  The branches use about eight $T$ gates per bit of precision, and the best lower bound is about one.  Algorithms such as Grover search call the oracle many times, so this constant multiplies their total $T$-count.  Is $\Tmix_\eps(\Toff_n)$ at least $\kappa\log(1/\eps)$ for some $\kappa>1$ when $n\ge\log(1/\eps)$?  For mixtures of oracle branches, \Cref{inf:oracle} reduces this to whether $\liminf_{k\to\infty}t(k)/k>1$, a question about exact multi-controlled NOTs.

Which property of the exceptional set $S$ determines the cost of $f=a\oplus\mathbf1_S$ with $a$ affine?  \Cref{inf:unstructured} settles the two extremes at constant error.  Subspaces cost $O(1)$, while random sets of density $2^{-k}$ cost about $2^{(n-k)/2}$ with clean ancillae, so distance from affine alone does not decide the cost.  Oracles in applications rarely sit at either extreme, and a compiler would want to know which of them are cheap without ancillae.  A natural candidate measure is the least number $J$ of affine subspaces whose indicators have exclusive-or $\mathbf1_S$.  \Cref{inf:affine} bounds the cost above by about $8J\ell_J$.  Is it also bounded below by a function of $J$?  Our counting argument gives lower bounds only for random sets, not for a given $S$.  It also fails with advice states, which can encode $S$, and we do not know whether random sets stay expensive with measurement and feedforward.

The branches matter most when logical qubits, rather than $T$ gates, limit a computation.  Magic-state cultivation may push computations toward that regime by making $T$ states much cheaper \cite{GidneyShuttyJones2024,Rosenfeld2025cultivation}, though its cost under realistic postselection is still open \cite{Mehta2026cultivation}, and early fault-tolerant devices will have few logical qubits in any case \cite{Katabarwa2024}.

Amplitude estimation without ancillae, for example, reflects about $\ket{0^n}$ on its whole register at every step, so no qubit is free to serve as an ancilla \cite{ErleKoczor2026}.  Compilers also treat clean qubits as scarce and borrow idle ones where they can \cite{SuZhouFengYing2025}.  GKZ's sampler, as written, computes its parities into clean ancillae, while the branches borrow only their own idle controls.  Being unitary, the branches also avoid the mid-circuit measurements and classically controlled corrections of the measured constructions in \Cref{tab:toffoli-resources}.

Like GKZ's sampler, the branches are accurate only on average, so the control system must draw a fresh branch for every call, using about $\ell n$ random bits.  Each branch applies $\Gamma$ and its inverse with $O(\ell n)$ CNOT gates, about as many as GKZ's direct circuit uses to compute its parities.  We have not estimated space-time volume on a specific architecture.

The branches cost more when $n$ is just above $\ell$.  For $\ell+2\le n<\lceil3\ell/2\rceil$ they can borrow only $d=n-1-\ell$ idle controls and use $16\ell-16d-16$ $T$ gates, up to twice the $8\ell-8$ they use with more idle controls.  We do not know how to reach $8\ell+O(1)$ with $O(1)$ idle controls.  The same question is open for exact multi-controlled NOTs with a few borrowed qubits, and an answer there would carry over to the branches directly.

\appendix

\section{Preliminaries}
\label{sec:prelim}

All logarithms are base $2$ unless written as $\ln$, and $[n]=\{1,\ldots,n\}$.  We identify a subset $S\subseteq[n]$ with its characteristic vector in $\F_2^n$, so $\abs S$ is also its Hamming weight.  We write $\mathbf1$ for the all-ones vector and $\mathbf1_A$ for the indicator function of a set $A$.

For $S,x\in\F_2^n$, the \emph{parity} $S\cdot x$ is the mod-$2$ sum of the bits of $x$ that $S$ selects.  The affine functions $\F_2^n\to\F_2$ are the maps $x\mapsto a\cdot x\oplus c$.

The oracle of a Boolean function $f:\F_2^n\to\F_2$ acts on $n$ input qubits and one target qubit as $U_f\ket{x}\ket{b}=\ket{x}\ket{b\oplus f(x)}$.  The Toffoli gate $\Toff_n$ on $n$ qubits is the oracle of the conjunction of $n-1$ bits, which are its controls.

\subsection{Distances and synthesis models}

For states $\rho,\sigma$, the trace distance is $D(\rho,\sigma)=\frac12\norm{\rho-\sigma}_1$, where $\norm{A}_1=\Tr\sqrt{A^\dagger A}$ is the trace norm.  We write $\norm{A}_\infty$ for the operator norm.  For channels $\mathcal E,\mathcal F$, the diamond distance is
\[
  D_\diamond(\mathcal E,\mathcal F)=\frac12\norm{\mathcal E-\mathcal F}_\diamond
  =\max_\rho\,D\bigl((\mathcal E\otimes I)(\rho),(\mathcal F\otimes I)(\rho)\bigr),
\]
where $\rho$ ranges over states of the channels' input system and a reference system on which both act as the identity.  We abbreviate $D_\diamond(\mathcal E,U)=D_\diamond(\mathcal E,\Phi_U)$, where $\Phi_U(\rho)=U\rho U^\dagger$.

\begin{definition}[mixed implementations]
\label{def:models}
A mixed implementation of $U$ is a probability distribution over unitary Clifford+$T$ circuits, called branches.  Each branch acts on the qubits of $U$ and on ancillae that start in $\ket0$ and are discarded at the end.  The error of the mixed implementation is the diamond distance of the averaged channel from $U$, and its cost is the largest number of $T$ gates on any branch.  The minimum cost at diamond error $\eps$ is $\Tmix_\eps(U)$.  Most lower bounds in \Cref{sec:lower} also hold when the ancillae start in an arbitrary advice state, as stated there.
\end{definition}

An ancilla is clean if it starts in $\ket0$ and must be returned to $\ket0$.  It is borrowed (often called dirty) if it may start in any state, even one entangled with the rest of the computation, and must be returned unchanged.  A branch with no ancilla acts only on the qubits of $U$, but it may borrow some of them as work qubits.

The next lemma, from GKZ \cite[Theorem~12]{gosset2025multiqubittoffoliexponentiallyfewer}, bounds the diamond error of a mixture by the largest probability, over all inputs, that a random branch computes that input wrongly.  We include its short proof.

\begin{lemma}[error from pointwise miss probabilities]
\label{lem:pointwise-miss}
Let $U$ be a unitary, and for each value of a random variable $a$ let $\Pi_a$ be a projector.  If
\[
  \left\|\E_a\Pi_a\right\|_\infty\le p,
\]
then the average of the channels $\Phi_{U(I-2\Pi_a)}$ is within diamond distance $4p$ of $U$.
In particular, let $f$ be Boolean and let $g_a$ be a random Boolean function.  If $\Prb_a[g_a(x)\ne f(x)]\le p$ for every $x$, then
\[
  D_\diamond\!\left(\E_a\Phi_{U_{g_a}},\Phi_{U_f}\right)\le4p.
\]
\end{lemma}

\begin{proof}
Each branch applies $I-2\Pi_a$ and then $U$, and the final $U$ does not change trace distances.  So it suffices to compare the average of $(I-2\Pi_a)\psi(I-2\Pi_a)$ with $\psi$.  By convexity, the maximum in the diamond distance may be taken over pure inputs, so let $\psi$ be a pure state of the system and the reference.  Write $\overline\Pi=\E_a\Pi_a$ for the averaged projector.  Expanding the product gives
\[
  \E_a[(I-2\Pi_a)\psi(I-2\Pi_a)]-\psi
  =-2\overline\Pi\psi-2\psi\overline\Pi+4\E_a[\Pi_a\psi\Pi_a].
\]
Since $\norm{AB}_1\le\norm{A}_\infty\norm{B}_1$ and $\psi$ has trace norm $1$, each of the first two terms has trace norm at most $2\norm{\overline\Pi}_\infty$.  The last term is positive semidefinite, so its trace norm equals its trace.  Since $\Pi_a^2=\Pi_a$, this trace is
\[
  4\E_a\Tr(\Pi_a\psi)=4\Tr(\overline\Pi\psi)
  \le4\norm{\overline\Pi}_\infty.
\]
The difference therefore has trace norm at most $8\norm{\overline\Pi}_\infty\le8p$, so the two states are at trace distance at most $4p$.  This holds for every pure input $\psi$, which proves the first statement.

For the second statement, take $U=U_f$ and let $\Pi_a$ project onto the inputs where $g_a$ and $f$ differ, with the target in $\ket-$:
\[
  \Pi_a=\sum_{x:g_a(x)\ne f(x)}\ket{x}\!\bra{x}\otimes\ket{-}\!\bra{-}.
\]
An oracle $U_g$ applies $X^{g(x)}$ to the target on input $x$.  Since $X\ket\pm=\pm\ket\pm$, it is diagonal in the basis $\ket{x}\ket{\pm}$: it fixes $\ket{x}\ket{+}$ and multiplies $\ket{x}\ket{-}$ by $(-1)^{g(x)}$.  Hence $U_f^\dagger U_{g_a}$ is diagonal too, with entry $-1$ on the range of $\Pi_a$ and $1$ elsewhere, so $U_{g_a}=U_f(I-2\Pi_a)$.  The averaged projector is diagonal, with entry $\Prb_a[g_a(x)\ne f(x)]\le p$ at $\ket{x}\ket{-}$ and $0$ elsewhere, so its operator norm is at most $p$.  The first statement then gives the second.
\end{proof}

\subsection{Multi-controlled NOTs with borrowed qubits}
\label{sec:mcx}

Write $C^mX$ for the $m$-controlled NOT, which flips its target when all $m$ controls are $1$.  The constructions in \Cref{sec:constructions} use three known circuits for it.

A relative-phase Toffoli gate is a Toffoli gate followed by a diagonal unitary on its three qubits.  Maslov gives one with $4$ $T$ gates, against $7$ for the Toffoli gate itself \cite[Fig.~3]{Maslov2016}.  Its inverse is the diagonal followed by the Toffoli gate.  So when a relative-phase Toffoli gate and its inverse enclose gates that touch its three qubits only as controls, the two diagonals cancel and the pair acts as two exact Toffoli gates.

\begin{figure}[!htb]
\centering
\begin{quantikz}[row sep={0.8cm,between origins}, column sep=0.45cm]
\lstick{$x$} & \ctrl{2} & & \ctrl{2} & & \\
\lstick{controls of $y$} & \qwbundle{} & \ctrl{1} & & \ctrl{1} & \\
\lstick{$b$ (borrowed)} & \ctrl{1} & \targ{} & \ctrl{1} & \targ{} & \rstick{$b$} \\
\lstick{target $t$} & \targ{} & & \targ{} & & \rstick{$t\oplus(x\wedge y)$}
\end{quantikz}
\caption{Toggle detection.  The target is toggled by $x\wedge b$ before and after $b$ is toggled by $y$, so it flips by $x\wedge b\oplus x\wedge(b\oplus y)=x\wedge y$ whatever $b$ held.  The last gate restores $b$.}
\label{fig:toggle}
\end{figure}
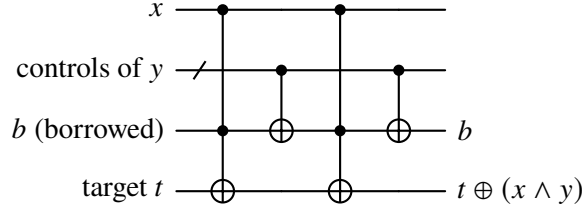

Borrowed qubits enter through the identity of \Cref{fig:toggle}.  A borrowed qubit $b$ holds an unknown bit.  The target is toggled by $x\wedge b$, then $b$ is toggled by the AND $y$ of other controls, and then the target is toggled by $x\wedge b$ again.  The two toggles of the target differ by $x\wedge y$, so the target flips by $x\wedge y$ whatever $b$ held, and a final toggle of $b$ by $y$ restores it.  Applied recursively along a chain, this builds $C^{m}X$ from $4(m-2)$ Toffoli gates and $m-2$ borrowed qubits \cite[Lemma~7.2]{barenco1995}.  Splitting the controls into two halves, each borrowing the other's qubits, brings this down to one borrowed qubit and $O(m)$ Toffoli gates \cite[Lemma~7.3 and Corollary~7.4]{barenco1995}.  Only the Toffoli gates that act on the target must be exact, and the others may be relative-phase gates.  Maslov also uses relative-phase gates with three controls, so that each borrowed qubit absorbs two controls.  For $m\ge4$ this gives $C^{m}X$ with $\lceil(m-2)/2\rceil$ borrowed qubits and $8m-8$ $T$ gates \cite[Proposition~5]{Maslov2016}.  For $m=3$ his circuit uses one borrowed qubit and $16$ $T$ gates \cite[Eq.~(5)]{Maslov2016}.

Zindorf and Bose use a borrowed qubit differently.  A rotation by $2\pi$ multiplies every state by $-1$, so a $2\pi$ rotation of the borrowed qubit, controlled by the $m$ controls and the target, is a multi-controlled $Z$ whatever the borrowed qubit holds.  Hadamard gates on the target turn it into $C^{m}X$.  For $m\ge5$ they implement it with $16(m+1)-48=16m-32$ $T$ gates, and each further borrowed qubit, up to $\lfloor(m-5)/2\rfloor$ of them, saves $16$ $T$ gates \cite[Theorems~2 and~10]{ZindorfBose2024}.

\section{Constructions without an ancilla}
\label{sec:constructions}

This appendix gives the constructions without an ancilla: the randomized Toffoli gate in \Cref{sec:toffoli} and its extension to exclusive-ors of affine subspaces in \Cref{sec:affine-xor}.  Both test a few random parities of the input with one multi-controlled NOT.  Inputs that the gate does not use serve as its borrowed work qubits.

\subsection{The randomized Toffoli gate}
\label{sec:toffoli}

\begin{construction}[the sampled branch]
\label{con:toffoli-branch}
Put $\ell=\lceil\log(4/\eps)\rceil$ and assume $\ell\le n-1$.  The case $\ell>n-1$ is the third part of \Cref{prop:toffoli-borrowed}.  Sample linearly independent $a_1,\ldots,a_\ell\in\F_2^{n-1}$ by drawing uniformly random tuples until one is independent.  Each draw succeeds with probability
\[
  \Prb[\text{independent}]=\prod_{i=0}^{\ell-1}(1-2^{i-n+1})
  \ge\prod_{j=1}^{\infty}(1-2^{-j})>0.28,
\]
so fewer than four draws are needed on average.  The branch then acts as follows.
\begin{enumerate}
\item Apply $X$ to all $n-1$ controls.
\item Row-reduce $a_1,\ldots,a_\ell$, which keeps their common kernel, and complete them to a basis $a_1,\ldots,a_{n-1}$ of $\F_2^{n-1}$ with unit vectors on the $n-1-\ell$ non-pivot coordinates.  Apply a CNOT circuit $\Gamma$ for the invertible linear map $\Gamma y=(a_1\cdot y,\ldots,a_{n-1}\cdot y)$, which needs $O(\ell n)$ CNOT gates.  We call the $n-1$ qubits that hold $\Gamma y$ the data coordinates.
\item Apply $C^{\ell}X$ with the first $\ell$ data coordinates as controls, active on $0$.
\item Apply $\Gamma^\dagger$ and then the $X$ layer again.
\end{enumerate}
On a basis state, with $y=x\oplus\mathbf1$ and $g(y)=\OR_{j\le\ell}(a_j\cdot y)$, the four steps give
\begin{align*}
  \ket{x}\ket{b}
  &\xmapsto{1}\ket{y}\ket{b}
  \xmapsto{2}\ket{\Gamma y}\ket{b}\\
  &\xmapsto{3}\ket{\Gamma y}\ket{b\oplus1\oplus g(y)}
  \xmapsto{4}\ket{x}\ket{b\oplus h(x)},
\end{align*}
where $h(x)=1\oplus g(x\oplus\mathbf1)$.  Step~3 flips the target exactly when every sampled parity of $y$ vanishes, so the branch is the oracle $U_h$.  It is GKZ's branch, drawn in \Cref{fig:toffoli}.  GKZ draw the $a_j$ uniformly, possibly dependent; the branch depends only on their span, so it can be computed in place either way, and we draw independent vectors only so that the counts below are exact.
\end{construction}

Since $\Gamma$ acts in place, no extra register holds the sampled parities.  The other $n-1-\ell$ data coordinates are idle during step~3, so $C^{\ell}X$ can borrow them as work qubits.

\begin{proposition}[determinants of the Toffoli gate and its branches]
\label{prop:determinant}
Let $n\ge4$.
\begin{enumerate}
\item No Clifford+$T$ circuit on the $n$ qubits of $\Toff_n$ implements it, even up to a global phase.
\item The oracle $U_h$ of a branch of \Cref{con:toffoli-branch} has determinant $1$ if $\ell\le n-2$ and $-1$ if $\ell=n-1$.
\end{enumerate}
\end{proposition}

\begin{proof}
Both parts rest on one fact: every Clifford+$T$ circuit on $N\ge4$ qubits has determinant $1$ \cite{GilesSelinger2013}.  The determinant is multiplicative, so it suffices to check each gate.  A gate $G$ on $j$ of the $N$ qubits acts on the whole register as $G\otimes I$, whose determinant is $\det(G)^{2^{N-j}}$.  The exponent is $2^{N-1}$ for a one-qubit gate, a multiple of $8$, and $2^{N-2}$ for CNOT, a multiple of $4$.  With $\omega=e^{i\pi/4}$, the generators have
\[
  \det T=\omega,\qquad \det S=i,\qquad \det H=-1,\qquad \det\mathrm{CNOT}=-1,
\]
and $\omega^8=i^8=(-1)^8=(-1)^4=1$.  So every gate, and hence every circuit, has determinant $1$.

For the first part, suppose a circuit implements $e^{i\theta}\Toff_n$.  The gate $\Toff_n$ swaps $\ket{1^{n-1},0}$ and $\ket{1^{n-1},1}$ and fixes all other basis states, so $\det\Toff_n=-1$ and the circuit's determinant is $-e^{i\theta2^n}$.  This must equal $1$, so $e^{i\theta}$ is a root of unity.  It is also the circuit's matrix entry at $\ket{0^n}$, which lies in $\mathbb Z[1/\sqrt2,i]\subset\mathbb Q(\omega)$ \cite{GilesSelinger2013}, and the only roots of unity in $\mathbb Q(\omega)$ are the powers of $\omega$.  Hence $e^{i\theta}=\omega^k$ and $e^{i\theta2^n}=\omega^{k2^n}=1$, since $8$ divides $2^n$, a contradiction.

For the second part, $U_h$ swaps $\ket{x,0}$ and $\ket{x,1}$ for each $x$ with $h(x)=1$ and fixes all other basis states.  It is a product of $\abs{h^{-1}(1)}$ disjoint transpositions, so $\det U_h=(-1)^{\abs{h^{-1}(1)}}$.  It remains to count these $x$.  By \Cref{con:toffoli-branch}, $h(x)=1$ exactly when $a_j\cdot y=0$ for $j=1,\ldots,\ell$, where $y=x\oplus\mathbf1$.  These are $\ell$ independent linear equations, so exactly $2^{n-1-\ell}$ vectors $y\in\F_2^{n-1}$ satisfy them.  Since $x\mapsto x\oplus\mathbf1$ is a bijection, $\abs{h^{-1}(1)}=2^{n-1-\ell}$.  This count is even when $\ell\le n-2$ and equals $1$ when $\ell=n-1$.
\end{proof}

The first part is the standard obstruction \cite{Shende2003,GilesSelinger2013,Maslov2016}.

\begin{remark}[which oracles need an ancilla]
\label{rem:degree-criterion}
The same computation applies to any oracle.  As in the second part, $\det U_f=(-1)^{\abs{f^{-1}(1)}}$.  In the algebraic normal form of $f$, the coefficient of a monomial $\prod_{i\in S}x_i$ is the exclusive-or of $f(x)$ over all $x$ whose ones lie in $S$ (M\"obius inversion).  For $S=[n]$ this is the exclusive-or of all values of $f$, so $\abs{f^{-1}(1)}$ is odd exactly when $\deg f=n$.  For $n\ge3$ the converse also holds, so $U_f$ has an exact Clifford+$T$ circuit on its $n+1$ qubits if and only if $\deg f<n$ \cite[Lemma~11, Theorem~12 and Corollary~37]{Shende2003}\cite[Corollary~2]{GilesSelinger2013}.
\end{remark}

\begin{theorem}[the randomized Toffoli gate]
\label{prop:toffoli-borrowed}
Let $0<\eps<1$ and $\ell=\lceil\log(4/\eps)\rceil$, so $\ell\ge3$.  If $\ell\le n-1$, the branches of \Cref{con:toffoli-branch} form a mixed implementation of $\Toff_n$ at diamond error $\eps$.  The cost depends on the number $d=n-1-\ell$ of idle data coordinates:
\begin{enumerate}
\item if $n\ge\lceil3\ell/2\rceil$, each branch has a circuit with $8\ell-8$ $T$ gates, no ancilla and no measurement;
\item if $\ell+2\le n<\lceil3\ell/2\rceil$, each branch has a circuit with $16\ell-16d-16\le16\ell-32$ $T$ gates, no ancilla and no measurement;
\item if $n\le\ell+1$, $\Toff_n$ has a mixed implementation at diamond error $\eps$ with one clean ancilla and $O(n)$ $T$ gates per branch.
\end{enumerate}
\end{theorem}

\Cref{fig:crossover} plots the three regimes at $\eps=10^{-6}$.

\begin{proof}[Proof of \Cref{prop:toffoli-borrowed}]
For the error, fix an input $x$ and let $y=x\oplus\mathbf1$.  By \Cref{con:toffoli-branch}, the branch flips the target exactly when $a_j\cdot y=0$ for every $j\le\ell$.  At $y=0$ this always holds, and $\Toff_n$ flips the target too, so the branch is correct.  At $y\ne0$ the gate $\Toff_n$ does not flip the target, so the branch errs exactly when every $a_j$ lies in the hyperplane $\{a:a\cdot y=0\}$, of dimension $n-2$.  A space of dimension $k$ contains $\prod_{i=0}^{\ell-1}(2^k-2^i)$ independent ordered $\ell$-tuples, since the $i$th vector must avoid the span of the earlier ones.  Comparing $k=n-2$ with $k=n-1$ gives
\[
  \Prb[a_j\cdot y=0\ \forall j]
  =\prod_{i=0}^{\ell-1}\frac{2^{n-2}-2^i}{2^{n-1}-2^i}
  \le 2^{-\ell},
\]
since $2^{n-2}-2^i\le\frac12(2^{n-1}-2^i)$.  So every input is computed wrongly with probability at most $2^{-\ell}$, and \Cref{lem:pointwise-miss} gives diamond error at most $4\cdot2^{-\ell}\le\eps$.

For the first part, the number of idle data coordinates is $d=n-1-\ell\ge\lceil3\ell/2\rceil-1-\ell=\lceil(\ell-2)/2\rceil$.  By \Cref{sec:mcx}, Maslov's circuit for $C^{\ell}X$ needs this many borrowed qubits and uses $8\ell-8$ $T$ gates, including $\ell=3$, where $16=8\ell-8$.  As noted after \Cref{con:toffoli-branch}, the idle coordinates can serve as these borrowed qubits.  The $X$ layers and $\Gamma$ are Clifford, so the branch uses $8\ell-8$ $T$ gates, no ancilla and no measurement.

For the second part, $n\ge\ell+2$ gives $d\ge1$, and $n<\lceil3\ell/2\rceil$ gives $d\le\lceil3\ell/2\rceil-2-\ell=\lfloor(\ell-3)/2\rfloor$.  In particular this range is nonempty only when $\ell\ge5$.  Zindorf and Bose's circuit for $C^{\ell}X$ (\Cref{sec:mcx}) can use the $d$ idle coordinates as borrowed qubits, since $d-1\le\lfloor(\ell-5)/2\rfloor$ by the bound above.  It uses
\[
  16(\ell+1)-16(d-1)-48=16\ell-16d-16
\]
$T$ gates and is unitary, so the branch needs no ancilla and no measurement.  For $d=1$ the count is $16\ell-32$, which Khattar and Gidney also give \cite[Table~1]{KhattarGidney2024}.

For the third part, suppose first that $n=\ell+1$.  No data coordinate is then idle, and for $n\ge4$ each branch has determinant $-1$ by \Cref{prop:determinant}, so it cannot be built on its own $n$ qubits.  We implement its gate $C^{\ell}X$ by Barenco et al.'s construction with one borrowed qubit (\Cref{sec:mcx}), which uses $O(\ell)=O(n)$ $T$ gates.  One clean ancilla serves as that qubit.  If $n\le\ell$, the same construction implements $\Toff_n$ exactly with one clean ancilla and $O(n)$ $T$ gates.
\end{proof}

\subsection{Affine decompositions}
\label{sec:affine-xor}

This subsection extends the branches from the Toffoli gate to exclusive-ors of indicators of affine subspaces.  Each term reduces to a multi-controlled NOT, so we first fix its cost.  Let $c(m)$ be the $T$-count of Maslov's circuit for $C^{m}X$ from \Cref{sec:mcx}, which borrows $\lceil(m-2)/2\rceil$ idle qubits.  For $m\le2$ the gate is a NOT, a CNOT or a Toffoli gate, so $c(0)=c(1)=0$ and $c(2)=7$, and $c(m)=8m-8$ for $m\ge3$.

\begin{theorem}[affine decompositions]
\label{prop:affine-xor}
Let $0<\eps<1$ and $f=\bigoplus_{j=1}^J\mathbf1_{A_j}$, where each $A_j\subseteq\F_2^n$ is an affine subspace cut out by $r_j$ independent affine equations.  Put $\ell_J=\lceil\log(4J/\eps)\rceil$.  If $n\ge\lceil3\min\{r_j,\ell_J\}/2\rceil-1$ for every $j$, then $U_f$ has a mixed Clifford+$T$ implementation at diamond error $\eps$ with no ancilla, no measurement and
\[
  \sum_{j=1}^Jc\bigl(\min\{r_j,\ell_J\}\bigr)\le 8J\ell_J
\]
$T$ gates per branch.  The terms with $r_j\le\ell_J$ are implemented exactly.
\end{theorem}

The hypothesis matches the first part of \Cref{prop:toffoli-borrowed}, shifted by one because $U_f$ has $n$ inputs while $\Toff_n$ has $n-1$ controls.  For $n\ge3$ it forces $\min\{r_j,\ell_J\}<n$.  So every exact or sampled term is the oracle of an affine subspace of codimension less than $n$, which has an even number of points.  By \Cref{rem:degree-criterion} each term then has determinant $1$, which is why no ancilla is needed even when $\deg f=n$.

\begin{proof}[Proof of \Cref{prop:affine-xor}]
Write $h_j=\mathbf1_{A_j}$.  Since $U_gU_h=U_{g\oplus h}$, we have $U_f=U_{h_J}\cdots U_{h_1}$, so it suffices to implement each $U_{h_j}$.  Fix $j$, put $r=r_j$, and write $A_j=\{x:Bx=c\}$ with $B$ of full row rank $r$.  As in \Cref{con:toffoli-branch}, complete the rows of $B$ to a basis in the same way and apply the CNOT circuit $\Gamma$ of the resulting invertible map, so that the first $r$ data coordinates hold $Bx$.  An $X$ gate on each coordinate $i$ with $c_i=1$ then makes $x\in A_j$ exactly when these $r$ coordinates are all $0$.

Suppose first that $r\le\ell_J$.  Then $C^{r}X$ on these $r$ coordinates, with controls active on $0$, flips the target exactly when $x\in A_j$.  Undoing the $X$ gates and $\Gamma$ gives $U_{h_j}$ exactly with $c(r)$ $T$ gates.  The circuit for $C^{r}X$ borrows $\lceil(r-2)/2\rceil$ of the other $n-r$ data coordinates, or none when $r\le2$.  There are enough, because the hypothesis $n\ge\lceil3r/2\rceil-1$ gives $n-r\ge\lceil r/2\rceil-1=\lceil(r-2)/2\rceil$.

Suppose next that $r>\ell_J$.  Write $z\in\F_2^r$ for the values of the $r$ coordinates, so that $x\in A_j$ exactly when $z=0$.  We run the branch of \Cref{con:toffoli-branch} on $z$, with $r$ in place of $n-1$ and $\ell_J$ in place of $\ell$.  It samples independent $s_1,\ldots,s_{\ell_J}\in\F_2^r$ and moves the parities $s_k\cdot z$ onto $\ell_J$ of the coordinates by a CNOT circuit.  It then applies $C^{\ell_J}X$ with controls active on $0$ and undoes the CNOT circuit, the $X$ gates and $\Gamma$.  The error computation in the proof of \Cref{prop:toffoli-borrowed} applies unchanged.  The branch is correct when $z=0$ and errs with probability at most $2^{-\ell_J}$ at each $z\ne0$.  By \Cref{lem:pointwise-miss}, the averaged term is within $4\cdot2^{-\ell_J}\le\eps/J$ of $U_{h_j}$.  Its gate $C^{\ell_J}X$ costs $c(\ell_J)$ $T$ gates and borrows $\lceil(\ell_J-2)/2\rceil$ of the other $n-\ell_J$ data coordinates.  There are enough, because $n\ge\lceil3\ell_J/2\rceil-1$ gives $n-\ell_J\ge\lceil(\ell_J-2)/2\rceil$.

Choose the randomness of the $J$ terms independently.  A branch for $U_f$ is then one branch per term, applied in sequence, and the averaged channel is the composition of the $J$ averaged term channels.  Each is within $\eps/J$ of its $U_{h_j}$.  The diamond distance is subadditive under composition, $D_\diamond(\mathcal E_2\mathcal E_1,\mathcal F_2\mathcal F_1)\le D_\diamond(\mathcal E_1,\mathcal F_1)+D_\diamond(\mathcal E_2,\mathcal F_2)$, so the averaged channel is within $\eps$ of $U_f$.  Since $c(m)\le8m$, every branch uses at most $\sum_jc(\min\{r_j,\ell_J\})\le8J\ell_J$ $T$ gates and makes no measurement.  It borrows only data coordinates and returns them unchanged, so it needs no ancilla.
\end{proof}

\subsubsection{Every Boolean function has such a decomposition}
\label{sec:general-decompositions}

Every Boolean function has such decompositions, and \Cref{prop:affine-xor} applies to each of them.
\begin{itemize}
\item The algebraic normal form writes $f$ as an exclusive-or of monomials $\prod_{i\in S}x_i$.  Each monomial is the indicator of the affine subspace $\{x:x_i=1\text{ for all }i\in S\}$, of codimension $\abs S$.
\item Any partition of $f^{-1}(1)$ into affine subspaces gives one, because the exclusive-or of the indicators of disjoint sets is the indicator of their union.  Single points give $J=\abs{f^{-1}(1)}$ terms of codimension $n$.
\end{itemize}
The decomposition should be chosen with care.  For example, the algebraic normal form of $\OR_n$ has all $2^n-1$ nonempty monomials, but $\OR_n=1\oplus\mathbf1_{\{0\}}$ is a single term of codimension $n$ followed by an $X$ gate on the target.

The pieces need not be disjoint.  The intersection of two affine subspaces is empty or an affine subspace, so $\mathbf1_{A\cup A'}=\mathbf1_A\oplus\mathbf1_{A'}\oplus\mathbf1_{A\cap A'}$ is again such a decomposition.  By \Cref{prop:affine-xor}, few terms of small codimension give a cheap oracle.  Finding the cheapest decomposition is the minimization of exclusive-or sums of affine products, for which reversible-logic synthesis has heuristics \cite{SchaefferPerkowski2013,FazelThorntonRice2007}.

A parity decision tree queries one parity of the input at each internal node and outputs a bit at each leaf.  A tree of depth $D$ with $J$ accepting leaves gives such a decomposition.  The inputs that reach a leaf satisfy the at most $D$ equations along its path, so each nonempty accepting leaf is an affine subspace of codimension at most $D$.  Distinct leaves are disjoint, so the second item above applies.  Since $c$ is nondecreasing, \Cref{prop:affine-xor} implements the oracle of the tree's function with at most $J\,c(\min\{D,\ell_J\})$ $T$ gates and no ancilla when $n\ge\lceil3\min\{D,\ell_J\}/2\rceil-1$.  With workspace, GKZ's bound in terms of randomized parity decision trees gives $O(J\min\{D,\ell_J\})$ \cite[Theorem~5]{gosset2025multiqubittoffoliexponentiallyfewer}.
\section{Lower bounds}
\label{sec:lower}

Every bound in this appendix holds for mixtures of unitary Clifford+$T$ circuits with any number of ancillae in any advice state.  The exceptions are \Cref{thm:oracle-equivalence}, \Cref{thm:random-exponential} and the bounds with clean ancillae or none in \Cref{cor:distance-not-enough}, which need ancillae that start in $\ket0$.

Most of the bounds go through randomized parity decision trees.  A non-adaptive randomized parity decision tree of depth $q$ is a parity decision tree, as in \Cref{sec:general-decompositions}, whose $q$ queries are drawn at random and fixed in advance.  Formally, it is a probability distribution over pairs consisting of $q$ parities of the input and a rule that maps their values to an output bit.  It computes $f$ with error $\eps$ if its output equals $f(x)$ with probability at least $1-\eps$ for every input $x$.  Write $\RPDT_\eps(f)$ for the least depth of such a tree.  These trees are randomized linear sketches over $\F_2$ \cite{KMSY18}, and the averaging arguments below are the standard ones from that setting \cite{KMSY18,MontanaroOsborne2009}.

GKZ show that $\Tmix_\eps(U_f)\ge\RPDT_\eps(f)-1$, even when the ancillae start in an arbitrary advice state \cite[Theorem~23 and its proof]{gosset2025multiqubittoffoliexponentiallyfewer}.  Girish, May, Parham and Yuen recover this reduction by a different technique \cite[Eq.~(21)]{GirishMayParhamYuen2025}.

\subsection{The Toffoli gate}
\label{sec:toffoli-lower}

\begin{proposition}[randomized parity decision trees for $\OR$]
\label{prop:or-rpdt}
For $m\ge1$, $0\le\eps<1$ and $0\le q\le m$, let $r_q=(2^{m-q}-1)/(2^m-1)$, the probability that a fixed nonzero vector lies in a uniformly random subspace of $\F_2^m$ of codimension $q$.  Then:
\begin{enumerate}
\item $\RPDT_\eps(\OR_m)=\min\{q:r_q\le\eps/(1-\eps)\}$;
\item GKZ's test of depth $\min\{q:r_q\le\eps\}$ computes $\OR_m$ with error $\eps$, and this depth exceeds the exact value by at most one when $\eps\le1/2$;
\item $\RPDT_\eps(\OR_m)\ge\min\{m,\log((1-\eps)/\eps)\}-1$.
\end{enumerate}
\end{proposition}

\begin{proof}[Proof of \Cref{prop:or-rpdt}]
For the second part, let $q$ satisfy $r_q\le\eps$.  The test of \Cref{con:toffoli-branch} chooses $q$ linearly independent parities uniformly at random and outputs $0$ exactly when all of them vanish.  Their common kernel $V$ is a uniformly random subspace of codimension $q$.  The output is correct at $x=0$, and at $x\ne0$ it is wrong exactly when $x\in V$.  Invertible linear maps send any nonzero vector to any other and leave the distribution of $V$ unchanged, so $\Prb[x\in V]$ is the same for every nonzero $x$.  Averaging over the $2^m-1$ nonzero vectors gives
\[
  \Prb[x\in V]=\frac{\abs{V\setminus\{0\}}}{2^m-1}=\frac{2^{m-q}-1}{2^m-1}=r_q\le\eps.
\]

For the first part, take a randomized tree of depth $q$ with error at most $\eps$.  We show that $r_q\le\eps/(1-\eps)$.  Let $\mu$ give weight $p=r_q/(1+r_q)$ to $0$ and spread the remaining weight $1-p$ uniformly over the nonzero inputs.  The tree errs with probability at most $\eps$ on every input, hence also on an input drawn from $\mu$.  That error is an average over the deterministic trees in its support, so one of them errs with $\mu$-probability at most $\eps$.  Its queries have a common kernel $V$ of codimension at most $q$.  Every input in $V$ gives the same answers, so the tree's output is constant on $V$.  If that output is $1$, the tree errs at $0$, with $\mu$-probability $p$.  If it is $0$, the tree errs on the at least $2^{m-q}-1$ nonzero inputs of $V$, with $\mu$-probability at least $(1-p)r_q$.  The choice of $p$ makes these equal,
\[
  p=(1-p)r_q=\frac{r_q}{1+r_q},
\]
so either way $\eps\ge p=r_q/(1+r_q)$, which rearranges to $r_q\le\eps/(1-\eps)$.  Since $r_q$ decreases in $q$, the depth is at least the least $q$ with this property.

To see that this depth is attained, let $q$ be least with $r_q\le\eps/(1-\eps)$ and put $\beta=r_q/(1+r_q)$, so that $\beta\le\eps$.  Consider the tree that outputs $1$ with probability $\beta$ and otherwise runs GKZ's test of depth $q$.  At $0$ it errs only when it outputs $1$, with probability $\beta$.  At a nonzero input it errs only when the test misses, with probability $(1-\beta)r_q=\beta$.  This tree has depth $q$ and error $\beta\le\eps$, which proves the first part.

For the gap in the second part, $2^{m-q-1}-1\le\frac12(2^{m-q}-1)$ gives $r_{q+1}\le r_q/2$, and $\eps\le1/2$ gives $\eps/(1-\eps)\le2\eps$.  So if $q$ is the exact value, then $r_{q+1}\le r_q/2\le\eps$, and GKZ's test needs depth at most $q+1$.

For the third part, put $\gamma=\eps/(1-\eps)$.  The condition $r_q\le\gamma$ is equivalent to $2^{m-q}\le1+(2^m-1)\gamma$, and the right side is at most $2\max\{1,2^m\gamma\}$.  Taking logarithms, every $q$ with $r_q\le\gamma$, in particular the exact value, satisfies
\[
  q\ge m-1-\max\{0,m+\log\gamma\}=\min\{m,\log(1/\gamma)\}-1.
\]
\end{proof}

Since $r_q\le2^{-q}$, the second part gives $\RPDT_\eps(\OR_m)\le\lceil\log(1/\eps)\rceil$.  So no lower bound for $\Toff_n$ obtained through the GKZ reduction can exceed $\lceil\log(1/\eps)\rceil-1$.

\begin{theorem}[lower bound for the Toffoli gate]
\label{prop:toffoli-lower}
For $n\ge2$ and $0<\eps\le1/2$, every mixed Clifford+$T$ implementation of $\Toff_n$ at diamond error $\eps$, with any number of ancillae in any advice state, has a branch with at least $\min\{n-1,\log((1-\eps)/\eps)\}-2$ $T$ gates.
\end{theorem}

\begin{proof}[Proof of \Cref{prop:toffoli-lower}]
The Toffoli gate is an $\OR$ oracle up to $X$ gates,
\[
  \Toff_n=(X^{\otimes(n-1)}\otimes I)\,U_{\OR_{n-1}}\,X^{\otimes n},
\]
since the target gains $1\oplus\OR_{n-1}(x\oplus\mathbf1)=x_1\wedge\cdots\wedge x_{n-1}$.  The $X$ gates are Clifford, so adding them turns a mixed implementation of either gate into one of the other with the same $T$-count and error.  The GKZ reduction and the third part of \Cref{prop:or-rpdt} with $m=n-1$ then give
\[
  \Tmix_\eps(\Toff_n)\ge\RPDT_\eps(\OR_{n-1})-1\ge\min\{n-1,\log((1-\eps)/\eps)\}-2.
\]
\end{proof}

\begin{corollary}[within a factor of eight]
\label{cor:factor-eight}
Let $0<\eps\le1/2$ and $\ell=\lceil\log(4/\eps)\rceil$, and suppose $n\ge\lceil3\ell/2\rceil$.  Let $T^*$ be the least branch $T$-count of any mixed implementation of $\Toff_n$ at diamond error $\eps$, with any ancillae in any advice state.  The branches of \Cref{prop:toffoli-borrowed} use no ancilla and at most $8T^*+40$ $T$ gates.
\end{corollary}

\begin{proof}[Proof of \Cref{cor:factor-eight}]
We compare the lower bound of \Cref{prop:toffoli-lower} with the count $8\ell-8$ of \Cref{prop:toffoli-borrowed}.  The hypothesis on $n$ gives
\[
  n-1\ge\lceil3\ell/2\rceil-1\ge\ell\ge\log(4/\eps)\ge\log((1-\eps)/\eps),
\]
so the minimum in \Cref{prop:toffoli-lower} is $\log((1-\eps)/\eps)$.  Using $1-\eps\ge1/2$, we get
\[
  T^*\ge\log((1-\eps)/\eps)-2\ge\log(1/\eps)-3.
\]
Since $\ell\le\log(4/\eps)+1$, the branches use
\[
  8\ell-8\le8(\log(4/\eps)+1)-8=8\log(1/\eps)+16\le8(T^*+3)+16=8T^*+40
\]
$T$ gates.
\end{proof}

For example, at $\eps=10^{-10}$ we have $\ell=36$, and the least $n$ with $n\ge\lceil3\ell/2\rceil$ is $54$.  There the GKZ reduction with the exact value from \Cref{prop:or-rpdt} gives a lower bound of $33$, while the branches use $280$ $T$ gates, a ratio of $8.5$.  The ratio tends to $8$ as $\eps\to0$.

\subsection{Oracle branches and exact multi-controlled NOTs}
\label{sec:oracle-branches}

The branches of \Cref{con:toffoli-branch}, like those of GKZ's sampler, each implement an exact oracle.  This subsection shows that for such mixtures the cost of $\Toff_n$ is pinned to the cost of exact multi-controlled NOTs.  Let $t(k)$ be the least $T$-count of a unitary Clifford+$T$ circuit that implements $C^{k}X$ exactly, with any number of clean ancillae and no measurement.  Known bounds give $k+1\le t(k)\le8k-9$ for $k\ge3$ \cite[Proposition~4.1]{Beverland2020}, \cite[Proposition~4]{Maslov2016}.

\begin{lemma}[steps of $t$]
\label{lem:t-steps}
For every $k\ge0$, $t(k)\le t(k+1)\le t(k)+8$.
\end{lemma}

\begin{proof}
For the first inequality, $C^{k}X$ is $C^{k+1}X$ with one control fixed to $1$.  So run an optimal circuit for $C^{k+1}X$ with that control on a clean ancilla, set to $\ket1$ by an $X$ gate before the circuit and reset afterwards.  This implements $C^{k}X$ with $t(k+1)$ $T$ gates.

For the second inequality, the case $k=0$ holds because $t(1)=0$, as $C^{1}X$ is a CNOT.  For $k\ge1$, merge two controls $a,b$ of $C^{k+1}X$ into one: write the AND of $a$ and $b$ into a clean ancilla $c$ with a relative-phase Toffoli gate, which uses $4$ $T$ gates (\Cref{sec:mcx}).  Then apply an optimal circuit for $C^{k}X$ with $c$ in place of $a$ and $b$, and undo the relative-phase gate with $4$ more $T$ gates.  In between, $C^{k}X$ touches $a,b,c$ only as controls, so the two relative-phase gates act as exact Toffoli gates and $c$ returns to $\ket0$.  The result implements $C^{k+1}X$ with $t(k)+8$ $T$ gates.
\end{proof}

A mixed implementation has \emph{oracle branches} if each branch implements an oracle $U_g$ exactly and returns its ancillae to $\ket0$, where the Boolean function $g$ may depend on the branch.  Write $\Tmix_{\eps,\mathrm{or}}$ for the least cost of such an implementation.

\begin{lemma}[pointwise errors of oracle branches]
\label{lem:oracle-pointwise}
If the oracle branches $U_{g_a}$ average to within diamond distance $\eps$ of $U_f$, then
$\Prb_a[g_a(x)\ne f(x)]\le\eps$ for every $x$.
\end{lemma}
\begin{proof}
Fix $x$.  Each branch returns its ancillae to $\ket0$, so we may ignore them.  Take a reference qubit $R$ and the input
\[
  \psi=\bigl(\ket{x}\ket{-}\ket0_R+\ket{x}\ket{+}\ket1_R\bigr)/\sqrt2.
\]
As in the proof of \Cref{lem:pointwise-miss}, an oracle $U_g$ multiplies $\ket{x}\ket{-}$ by $(-1)^{g(x)}$ and fixes $\ket{x}\ket{+}$, so the branch $U_{g_a}$ maps $\psi$ to
\[
  \psi_a=\bigl((-1)^{g_a(x)}\ket{x,-,0}+\ket{x,+,1}\bigr)/\sqrt2.
\]
Write $\psi_f$ for the ideal output and $\rho=\E_a\psi_a\psi_a^\dagger$ for the averaged output.  If $g_a(x)=f(x)$ then $\psi_a=\psi_f$.  Otherwise the two differ by a relative sign and are orthogonal.  Hence $\langle\psi_f|\rho|\psi_f\rangle=\Prb_a[g_a(x)=f(x)]$.  For any projector $P$, $D(\sigma,\rho)\ge\Tr P\sigma-\Tr P\rho$, and taking $P=\psi_f\psi_f^\dagger$ gives
\[
  \eps\ge D(\psi_f\psi_f^\dagger,\rho)\ge1-\langle\psi_f|\rho|\psi_f\rangle=\Prb_a[g_a(x)\ne f(x)].
\]
\end{proof}

Together, \Cref{lem:oracle-pointwise,lem:pointwise-miss} show that for oracle branches the diamond error is the largest pointwise error up to a factor of $4$:
\[
  \max_x\Prb_a[g_a(x)\ne f(x)]\le D_\diamond\le4\max_x\Prb_a[g_a(x)\ne f(x)].
\]

\begin{theorem}[oracle branches and exact multi-controlled NOTs]
\label{thm:oracle-equivalence}
Let $0<\eps\le1/2$ and $n-1\ge\lceil\log(4/\eps)\rceil$.  Then
\[
  t\bigl(\lfloor\log((1-\eps)/\eps)\rfloor\bigr)\;\le\;\Tmix_{\eps,\mathrm{or}}(\Toff_n)\;\le\;
  t\bigl(\lceil\log(4/\eps)\rceil\bigr).
\]
With $K=\lfloor\log(1/\eps)\rfloor$, this gives $t(K)-8\le\Tmix_{\eps,\mathrm{or}}(\Toff_n)\le t(K)+24$,
so $\Tmix_{\eps,\mathrm{or}}(\Toff_n)/\log(1/\eps)$ has the same $\liminf$ and $\limsup$ as
$t(k)/k$ when $\eps\to0$ with $n\ge\log(4/\eps)+1$.
\end{theorem}

\begin{proof}
As in the proof of \Cref{prop:toffoli-lower}, conjugating by $X$ gates replaces $\Toff_n$ by $U_f$ with $f=\OR_m$ and $m=n-1$.  This keeps oracle branches oracle branches and does not change $T$-counts.

For the upper bound, put $\ell=\lceil\log(4/\eps)\rceil\le m$ and take GKZ's sampler \cite[Theorem~1]{gosset2025multiqubittoffoliexponentiallyfewer}, that is, the branch of \Cref{con:toffoli-branch}, with its gate $C^{\ell}X$ implemented by an optimal exact circuit with clean ancillae.  Each branch is an oracle and uses $t(\ell)$ $T$ gates.  By the error computation in the proof of \Cref{prop:toffoli-borrowed}, every input is wrong with probability at most $2^{-\ell}$, so \Cref{lem:pointwise-miss} gives error at most $4\cdot2^{-\ell}\le\eps$.

For the lower bound, take oracle branches $U_{g_a}$ with error $\eps$ and put $\gamma=\eps/(1-\eps)$.  We find a branch that computes $\OR$ exactly on a subspace of dimension about $\log(1/\gamma)$, and turn it into an exact $C^{w}X$.  Let $Z_a=g_a^{-1}(0)$ be the set of inputs that branch $a$ rejects.  The ideal $f$ rejects only $0$, so \Cref{lem:oracle-pointwise} at $0$ and summed over the nonzero inputs gives
\[
  \Prb_a[0\notin Z_a]\le\eps
  \qquad\text{and}\qquad
  \E_a\abs{Z_a\setminus\{0\}}\le\eps(2^m-1).
\]
Since $\abs{Z_a\setminus\{0\}}\ge0$, conditioning on the event $0\in Z_a$, which has probability at least $1-\eps$, raises its expectation by at most a factor $1/(1-\eps)$.  Some branch therefore has
\[
  0\in Z_a
  \qquad\text{and}\qquad
  \abs{Z_a\setminus\{0\}}\le\gamma(2^m-1).
\]

We now look for a subspace $W$ of dimension $w=\lfloor\log(1/\gamma)\rfloor$ that avoids every nonzero input the branch rejects, that is, with $W\cap Z_a=\{0\}$.  On such a $W$ the branch rejects only $0$, so in coordinates of $W$ it computes $\OR_w$.  Note that $w\le m$, since $m\ge\log(4/\eps)>\log(1/\gamma)$.

To find $W$, choose it uniformly at random among subspaces of dimension $w$.  By the symmetry argument in the proof of \Cref{prop:or-rpdt}, each nonzero vector lies in $W$ with probability $(2^w-1)/(2^m-1)$.  The set $Z_a$ has at most $\gamma(2^m-1)$ nonzero vectors, so the expected number of them in $W$ is at most $\gamma(2^w-1)<\gamma2^w\le1$, where the last step is the choice of $w$.  A count whose expectation is below $1$ is $0$ for some choice, so some $W$ contains no nonzero vector of $Z_a$.  Since $0\in Z_a$, this $W$ has $W\cap Z_a=\{0\}$.

It remains to turn this branch into an exact circuit for $C^{w}X$.  Let $M$ be an invertible linear map with $M(\F_2^w\times\{0\})=W$.  On $w$ controls $z$, a target and $m-w$ clean ancillae, apply the CNOT circuit of $M$, then the branch, then the CNOT circuit of $M^{-1}$.  The input $(z,0^{m-w})$ becomes $M(z,0)\in W$.  The branch adds $g_a(M(z,0))=\OR_w(z)$ to the target and leaves the input register unchanged, and $M^{-1}$ restores $(z,0^{m-w})$.  So the circuit implements $U_{\OR_w}$ exactly, returns its ancillae to $\ket0$, and uses the branch's $T$ gates, which are at most the cost of the mixture.  With $X$ gates as in the proof of \Cref{prop:toffoli-lower} it becomes $C^{w}X$.  Hence $t(w)\le\Tmix_{\eps,\mathrm{or}}(\Toff_n)$ with $w=\lfloor\log(1/\gamma)\rfloor=\lfloor\log((1-\eps)/\eps)\rfloor$.

For the second statement, $\eps\le1/2$ gives
\[
  \lfloor\log((1-\eps)/\eps)\rfloor\ge K-1
  \qquad\text{and}\qquad
  \lceil\log(4/\eps)\rceil\le K+3.
\]
Since $t$ is nondecreasing, the first statement and \Cref{lem:t-steps} give
\[
  t(K)-8\le t(K-1)\le\Tmix_{\eps,\mathrm{or}}(\Toff_n)\le t(K+3)\le t(K)+24.
\]
As $\eps\to0$, $K$ runs through all large integers and differs from $\log(1/\eps)$ by less than one.  Dividing by $\log(1/\eps)$ therefore gives the statement about $\liminf$ and $\limsup$.
\end{proof}

\subsection{The distance from affine}
\label{sec:distance}

The bounds so far concern the Toffoli gate.  This subsection bounds every oracle in terms of one number, its distance from the nearest affine function.  For an affine $a$, write $\delta=\Prb_x[f(x)\ne a(x)]$ for the fraction of inputs on which $f$ and $a$ differ.  The distance from affine $\delta_H(f)$ is the least such $\delta$.  It is at most $1/2$, because $a$ and $a\oplus1$ are both affine and one of them agrees with $f$ on at least half of the inputs.

\begin{proposition}[decision trees and the distance from affine]
\label{prop:distance-rpdt}
Let $f$ be non-affine, let $a$ be affine with $\delta=\Prb_x[f(x)\ne a(x)]\le1/2$, and let $0<\eps\le1/2$.  Then
\[
  \RPDT_\eps(f)\ge\min\{\log(1/\delta),\log(1/\eps)-1\}-2.
\]
\end{proposition}

\begin{proof}[Proof of \Cref{prop:distance-rpdt}]
We reduce to the function $e=f\oplus a$, which is $1$ only on the small set $E=\{x:e(x)=1\}$, of size $\abs E=\delta2^n$.  Since $f$ is not affine, $e$ is neither $0$ nor $1$, so both $E$ and its complement are nonempty.  A randomized tree for $f$ of depth $q$, plus one query of the parity that defines $a$, is a randomized tree for $e$ of depth $q+1$ with the same error.  Hence
\[
  \RPDT_\eps(e)\le\RPDT_\eps(f)+1,
\]
and it suffices to show that every randomized tree for $e$ with error at most $\eps$ has depth $q\ge\min\{\log(1/\delta),\log(1/\eps)-1\}-1$.

Take such a tree, and let $\mu$ put weight $1/2$ uniformly on $E$ and weight $1/2$ uniformly on its complement.  As in the proof of \Cref{prop:or-rpdt}, some deterministic tree in its support errs with $\mu$-probability at most $\eps$.  Its queries have a common kernel $V$ of codimension at most $q$, so its output is constant on each coset $C$ of $V$.  These cosets are large but meet $E$ in few points, so the tree cannot isolate $E$:
\begin{enumerate}
\item If it outputs $1$ on some coset $C$, it errs on the at least $2^{n-q}-\delta2^n$ points of $C\setminus E$, each of $\mu$-weight at least $2^{-(n+1)}$.  Its error is then at least
\[
  \alpha=\frac{2^{-q}-\delta}{2}.
\]
\item If it outputs $0$ everywhere, it errs on all of $E$, with $\mu$-probability $1/2\ge\alpha$.
\end{enumerate}
Either way $\eps\ge\alpha$, which rearranges to
\[
  2^{-q}\le2\eps+\delta\le2\max\{2\eps,\delta\},
  \qquad\text{that is}\qquad
  q\ge\min\{\log(1/\delta),\log(1/\eps)-1\}-1.
\]
Combining this with $\RPDT_\eps(e)\le\RPDT_\eps(f)+1$ proves the proposition.
\end{proof}

\begin{theorem}[distance from affine]
\label{thm:distance}
For every non-affine $f$ and $0<\eps\le1/2$, every mixture of unitary Clifford+$T$ circuits that approximates $U_f$ to diamond error $\eps$, with any number of ancillae in any advice state, has a branch with at least $\min\{\log(1/\delta_H(f)),\log(1/\eps)-1\}-3$ $T$ gates.
\end{theorem}

\begin{proof}[Proof of \Cref{thm:distance}]
Take an affine $a$ attaining $\delta_H(f)$, which is at most $1/2$.  The GKZ reduction and \Cref{prop:distance-rpdt} give
\[
  \Tmix_\eps(U_f)\ge\RPDT_\eps(f)-1\ge\min\{\log(1/\delta_H(f)),\log(1/\eps)-1\}-3.
\]
\end{proof}

For the Toffoli gate, the $\AND$ of $n-1$ bits differs from the constant $0$ only at the all-ones input, so $\delta_H=2^{1-n}$ and \Cref{thm:distance} recovers \Cref{prop:toffoli-lower} up to an additive constant.  The bound is tight up to a factor of eight whenever $f$ differs from an affine function on a single affine subspace, as the next corollary shows.

\begin{corollary}[one exceptional subspace]
\label{cor:distance-two-sided}
Let $f=a\oplus\mathbf1_A$, where $a$ is affine and $A$ is an affine subspace of codimension $r\ge2$.  Let $0<\eps\le1/2$ and $\ell=\lceil\log(4/\eps)\rceil$, and let $T^*$ be the least branch $T$-count of any mixture of unitary circuits that approximates $U_f$ to diamond error $\eps$, with any ancillae in any advice state.  Then:
\begin{enumerate}
\item $\min\{r,\log(1/\eps)-1\}-3\le T^*\le8\min\{r,\ell\}-8$;
\item when $n\ge\lceil3\min\{r,\ell\}/2\rceil-1$, the upper bound is attained with no ancilla and no measurement, and this construction uses at most $8T^*+48$ $T$ gates;
\item $T^*=\Theta(\log(1/\max\{2^{-r},\eps\}))$ as $\min\{r,\log(1/\eps)\}\to\infty$.
\end{enumerate}
\end{corollary}

\begin{proof}[Proof of \Cref{cor:distance-two-sided}]
For the lower bound in the first part, $f$ differs from $a$ exactly on $A$, so $\delta_H(f)\le2^{-r}$.  Also $f$ is not affine: two distinct affine functions differ on half of the inputs or on all of them, while $f$ differs from $a$ on a fraction $2^{-r}\le1/4$.  \Cref{thm:distance} then gives the lower bound, since $\log(1/\delta_H(f))\ge r$.

For the upper bound and the second part, $U_f=U_aU_{\mathbf1_A}$, where $U_a$ is a CNOT circuit, possibly with an $X$ on the target.  It is Clifford, so only $U_{\mathbf1_A}$ costs $T$ gates.  When $n\ge\lceil3\min\{r,\ell\}/2\rceil-1$, \Cref{prop:affine-xor} with $J=1$ implements $U_{\mathbf1_A}$ with no ancilla and no measurement, using
\[
  c(\min\{r,\ell\})\le8\min\{r,\ell\}-8
\]
$T$ gates, since $c(m)\le8m-8$ for $m\ge2$.  For smaller $n$ the same circuits work with $\lceil(\min\{r,\ell\}-2)/2\rceil$ clean ancillae as their borrowed qubits, which $T^*$ allows.

For the factor of eight in the second part, write $L=\min\{r,\log(1/\eps)-1\}-3$ for the lower bound.  Since $\ell\le\log(1/\eps)+3$,
\[
  \min\{r,\ell\}\le\min\{r,\log(1/\eps)+3\}\le L+7,
\]
and the construction uses at most $8(L+7)-8=8L+48\le8T^*+48$ $T$ gates.  For the third part, both sides of the first part are $\Theta(\min\{r,\log(1/\eps)\})$, and $\min\{r,\log(1/\eps)\}=\log(1/\max\{2^{-r},\eps\})$.
\end{proof}

Toffoli gates, membership oracles of linear codes and all their affine shifts have this form.  Each such $U_f$ equals $C^{r}X$ up to Clifford gates, so an exact implementation needs at least $r+1$ $T$ gates \cite{Beverland2020}, while the exact constructions of \Cref{sec:related} use about $8r$.

\subsection{Random exceptional sets}
\label{sec:unstructured}

Two oracles at the same distance from affine can have very different costs.  At constant error, the indicator of an affine subspace of codimension $k$ is at distance about $2^{-k}$ from affine and costs $O(1)$ $T$ gates (\Cref{cor:distance-two-sided}).  A random set $R$ of the same density is at the same distance, but we show that it costs far more, in two steps.
\begin{enumerate}
\item \emph{Decision trees.}  A random set meets every coset of every subspace of small codimension in about its expected proportion.  A shallow tree, whose output is constant on such cosets, then cannot separate $R$ from its complement, so it needs depth close to $n-k$ (\Cref{thm:random-exceptional}).  This adapts the affine-extractor argument of Kannan et al.\ \cite[Lemmas~B.2 and~B.6]{KMSY18}.  As in \Cref{prop:distance-rpdt}, giving $R$ half the weight keeps the error constant at every density.
\item \emph{Counting.}  Circuits with few $T$ gates have few possible acceptance probabilities, too few to single out most sets of density $2^{-k}$.  So random sets need exponentially many $T$ gates (\Cref{thm:random-exponential}), far more than decision trees can certify.
\end{enumerate}
For $R\subseteq\F_2^n$ with $R$ and its complement nonempty, let $\mu_R$ give weight $1/2$ uniformly to $R$ and weight $1/2$ uniformly to its complement.

\begin{lemma}[equidistribution on cosets]
\label{lem:coset-discrepancy}
Let $V\subseteq\F_2^n$ be a subspace of codimension $q$, let $0<p\le1/2$ and $0\le\eta<1$, and suppose $\bigl|\abs{C\cap R}-p\abs C\bigr|\le\eta p\abs C$ for every coset $C$ of $V$.  Then every rule that is constant on the cosets of $V$ disagrees with $\mathbf1_R$ with $\mu_R$-probability at least $\frac12-\frac{\eta}{1-\eta^2}$.
\end{lemma}

\begin{proof}
Each coset receives nearly the same share of $R$ as of its complement, so a rule that is constant on cosets cannot tell the two halves of $\mu_R$ apart.

First, $\mu_R$ is well defined.  Summing the hypothesis over the $2^q$ cosets gives $\abs R\in[1-\eta,1+\eta]\,p2^n$, so $R$ is nonempty.  Each coset has $\abs{C\setminus R}\ge(1-p-\eta p)\abs C>0$, so the complement of $R$ is nonempty too.

Let
\[
  a_C=\frac{\abs{C\cap R}}{\abs R}
  \qquad\text{and}\qquad
  b_C=\frac{\abs{C\setminus R}}{\abs{\F_2^n\setminus R}}
\]
be the fractions of $R$ and of its complement that lie in $C$.  The hypothesis and the bound on $\abs R$ give $a_C\in2^{-q}[\frac{1-\eta}{1+\eta},\frac{1+\eta}{1-\eta}]$.  The same holds for $b_C$, because $\abs{C\setminus R}$ and $\abs{\F_2^n\setminus R}$ are within relative error $\eta p/(1-p)\le\eta$ of $(1-p)\abs C$ and $(1-p)2^n$.  Both lie in the same interval of length $2^{-q}\cdot4\eta/(1-\eta^2)$, and hence
\[
  \abs{a_C-b_C}\le2^{-q}\cdot\frac{4\eta}{1-\eta^2}.
\]

On a coset $C$ the rule errs with $\mu_R$-probability $a_C/2$ if it outputs $0$ there and $b_C/2$ if it outputs $1$.  Using $\sum_Ca_C=\sum_Cb_C=1$ and summing over the $2^q$ cosets, the rule errs with probability at least
\[
  \tfrac12\textstyle\sum_C\min\{a_C,b_C\}=\tfrac12\Bigl(1-\tfrac12\sum_C\abs{a_C-b_C}\Bigr)\ge\tfrac12-\tfrac{\eta}{1-\eta^2}.
\]
\end{proof}

\begin{lemma}[from cosets to decision trees]
\label{lem:coset-to-rpdt}
If the hypothesis of \Cref{lem:coset-discrepancy} holds with $\eta\le1/5$ for every subspace of codimension exactly $q\le n$, then $\RPDT_\eps(\mathbf1_R)\ge q+1$ and $\Tmix_\eps(U_{\mathbf1_R})\ge q$ for every $\eps\le1/4$, with any ancillae in any advice state.
\end{lemma}

\begin{proof}
Suppose a randomized tree of depth at most $q$ has error at most $\eps$.  Averaging over $\mu_R$, some deterministic tree in its support errs with $\mu_R$-probability at most $\eps$.  Its queries have a common kernel $V'$ of codimension at most $q$, and its output is constant on the cosets of $V'$.  It is then also constant on the cosets of any subspace $V\subseteq V'$ of codimension exactly $q$.  The bound of \Cref{lem:coset-discrepancy} decreases in $\eta$, so applying it to $V$ with $\eta\le1/5$ bounds the error below by
\[
  \frac12-\frac{1/5}{1-1/25}=\frac12-\frac5{24}=\frac7{24}>\frac14\ge\eps,
\]
a contradiction.  Hence $\RPDT_\eps(\mathbf1_R)\ge q+1$, and the GKZ reduction gives $\Tmix_\eps(U_{\mathbf1_R})\ge q$.
\end{proof}

\begin{theorem}[random exceptional sets]
\label{thm:random-exceptional}
Let $n\ge2$ and $1\le k\le n$, and let $s\ge0$ be an integer with $q=n-k-s\ge0$ and
\begin{equation}\label{eq:s-condition}
  2^s\ge52\bigl(q(k+s+1)+n+3\bigr).
\end{equation}
Let $R\subseteq\F_2^n$ contain each point independently with probability $2^{-k}$, or be a uniformly random set of size $2^{n-k}$.  With probability at least $1-2^{-n}$, every coset $C$ of every subspace of codimension at most $q$ satisfies $\abs{C\cap R}=(1\pm\frac15)2^{-k}\abs C$.  On this event, for every $\eps\le1/4$,
\[
  \RPDT_\eps(\mathbf1_R)\ge n-k-s+1
  \qquad\text{and}\qquad
  \Tmix_\eps(U_{\mathbf1_R})\ge n-k-s,
\]
the latter for mixtures of unitary Clifford+$T$ circuits with any ancillae in any advice state.  If moreover $k\ge2$ and $q\ge1$, then also $\delta_H(\mathbf1_R)=\abs R/2^n\in[\frac45,\frac65]\,2^{-k}$.  Condition~\eqref{eq:s-condition} holds for $s=\lceil2\log(n+1)\rceil+6$, and also for some $s=\log n+\log(k+\log n)+O(1)$.
\end{theorem}

\begin{proof}
Fix a subspace $V$ of codimension exactly $q$ and a coset $C$ of $V$.  Then $\abs C=2^{n-q}$, and $X=\abs{C\cap R}$ has mean $2^{n-q-k}=2^s$.  For independent inclusion, $X$ is binomial.  For a uniformly random set of fixed size it is hypergeometric, which satisfies the same Chernoff bounds \cite[Theorem~4]{Hoeffding1963}.  In both cases
\[
  \Prb\bigl[\abs{X-2^s}>2^s/5\bigr]\le2e^{-2^s/75}.
\]

The number of subspaces of codimension $q$ is a Gaussian binomial coefficient, at most $2^{q(n-q)}\prod_{i\ge1}(1-2^{-i})^{-1}<4\cdot2^{q(n-q)}$, and each has $2^q$ cosets.  The probability that some coset fails is therefore at most
\[
  8\cdot2^{q(n-q+1)}e^{-2^s/75}\le2^{-n}.
\]
The last inequality is \eqref{eq:s-condition} after taking logarithms, since $n-q=k+s$ and $75\ln2<52$.  Every coset of a subspace of codimension $q'<q$ is a disjoint union of cosets of a subspace of codimension $q$ contained in it.  The estimate therefore extends to all codimensions at most $q$.  \Cref{lem:coset-to-rpdt} with $p=2^{-k}$ and $\eta=1/5$ gives the bounds on $\RPDT$ and $\Tmix$.

For the distance, the case $q'=0$ gives $\abs R/2^n\in[\frac45,\frac65]2^{-k}$, which is the distance to the constant $0$.  A non-constant affine $a$ splits $\F_2^n$ into the hyperplane cosets $H_0=a^{-1}(0)$ and $H_1=a^{-1}(1)$.  The case $q'=1$ gives $\abs{R\cap H_0}-\abs{R\cap H_1}\ge-\frac15 2^{-k}2^n$.  When $k\ge2$, it follows that
\[
  \Prb_x[\mathbf1_R(x)\ne a(x)]=\frac12+\frac{\abs{R\cap H_0}-\abs{R\cap H_1}}{2^n}\ge\frac12-\frac1{20}>\frac65\cdot\frac14,
\]
so $a$ is farther from $\mathbf1_R$ than the constant $0$.  The constant $1$ is farther still.

For the choices of $s$, since $q+(k+s+1)=n+1$, we have $q(k+s+1)\le(n+1)^2/4$.  For $s=\lceil2\log(n+1)\rceil+6$ and $n\ge2$,
\[
  2^s\ge64(n+1)^2\ge13(n+1)^2+52(n+3)\ge52\bigl(q(k+s+1)+n+3\bigr).
\]
Since $q\le n$, \eqref{eq:s-condition} also holds once $2^s\ge52(n(k+s+1)+n+3)$, which is satisfied by some $s=\log n+\log(k+\log n)+O(1)$.
\end{proof}

Like the canonical-form count of Gosset, Kothari and Wu \cite[Lemma~4.8]{GKW24}, the first bound in the following count does not depend on the number of ancillae.

\begin{lemma}[few acceptance functions]
\label{lem:acceptance-count}
Let $U$ be a unitary Clifford+$T$ circuit with at most $t$ $T$ gates on $n$ input qubits, a target qubit and $a$ ancillae, and let $p_U(x)$ be the probability that measuring the target of $U\ket{x}\ket0\ket{0^a}$ gives $1$.  As $U$ ranges over all such circuits, there are at most $2^{(t+1)\min\{2t+n+3,\,2(n+a)+3\}}$ distinct functions $p_U:\F_2^n\to[0,1]$.
\end{lemma}

\begin{proof}
The function $p_U$ depends on $U$ only through a little combinatorial data about the Pauli rotations in $U$, and we count that data.  Up to a global phase, $U=C\,R(P_t)\cdots R(P_1)$ with a Clifford $C$, Hermitian Paulis $P_j$ and $R(P)=\exp(-i\pi P/8)$ \cite{GKMR14}, \cite[proof of Lemma~22]{gosset2025multiqubittoffoliexponentiallyfewer}.  A circuit with fewer $T$ gates is padded with $P_j=I$.  Let $\Pi$ project the target onto $\ket1$, put $Q_0=C^\dagger Z_{\mathrm{tgt}}C$, a Hermitian Pauli, and let $y=(x,0,0^a)$.  Then $p_U(x)=\bra yU^\dagger\Pi U\ket y=(1-\bra yA\ket y)/2$ with $A=R(P_1)^\dagger\cdots R(P_t)^\dagger Q_0R(P_t)\cdots R(P_1)$.

Conjugation by $R(P)$ fixes a Pauli $M$ that commutes with $P$ and sends one that anticommutes with $P$ to $M(I-iP)/\sqrt2$.  Expanding term by term, $A=\sum_Sc_SM_S$ over $S\subseteq\{1,\ldots,t\}$, where $M_S$ is the product of $Q_0$ and the $P_j$ with $j\in S$ in decreasing order of $j$, and each $c_S$ depends only on which pairs among $F_0=Q_0,F_1=P_1,\ldots,F_t=P_t$ anticommute.  Write $F_i=\sigma_iX(v_i)Z(w_i)$ with $\sigma_i\in\{\pm1,\pm i\}$ and $v_i,w_i\in\F_2^{n+1+a}$.  Since $X(v)Z(w)X(v')Z(w')=(-1)^{w\cdot v'}X(v+v')Z(w+w')$, each $M_S$ is $\pm\prod_{i\in S\cup\{0\}}\sigma_i$ times $X(v_S)Z(w_S)$, with the sign fixed by the bits $w_i\cdot v_j$, $v_S=\sum_{i\in S\cup\{0\}}v_i$ and $w_S=\sum_{i\in S\cup\{0\}}w_i$.  On the basis state $y$, $\bra yX(v)Z(w)\ket y$ is $(-1)^{w\cdot y}$ if $v=0$ and $0$ otherwise, and $w\cdot y$ depends only on the coordinates of $w$ on the input qubits.

Hence $p_U$ is determined by four pieces of data:
\begin{enumerate}[nosep]
\item the phases $\sigma_i$, with $4^{t+1}$ choices,
\item the bits $w_i\cdot v_j$ for $i\ne j$, which also fix which pairs anticommute, with $2^{t(t+1)}$ choices,
\item the subspace of indicator vectors $S'\in\F_2^{t+1}$ with $\sum_{i\in S'}v_i=0$, with at most $2^{(t+1)^2}$ choices,
\item the input coordinates of the $w_i$, with $2^{n(t+1)}$ choices.
\end{enumerate}  The product is $2^{(t+1)(2t+n+3)}$.  Alternatively, $p_U$ is determined by $Q_0,P_1,\ldots,P_t$ themselves, each one of $2\cdot4^{n+1+a}$ Hermitian Paulis, which gives $2^{(t+1)(2(n+a)+3)}$.
\end{proof}

\begin{theorem}[counting bound for random sets]
\label{thm:random-exponential}
Let $n\ge2$, $1\le k\le n$ and $0\le\eps<1/2$, put $N=\lceil(n+1)/(2(1/2-\eps)^2)\rceil$, and let $R$ be random as in \Cref{thm:random-exceptional}.  With probability at least $1-2^{1-n}$, every mixture of unitary Clifford+$T$ circuits with clean ancillae that approximates $U_{\mathbf1_R}$ to diamond error $\eps$ has cost $t$ with
\[
  N(t+1)(2t+n+3)\ge2^{n-k}-n,
\]
and every such mixture without ancillae has cost $t$ with $N(t+1)(2n+3)\ge2^{n-k}-n$.  At $\eps=1/4$, where $N=8(n+1)$, this gives
\[
  t\ge\sqrt{\frac{2^{n-k}-n}{16(n+1)}}-n-1
  \qquad\text{and}\qquad
  t\ge\frac{2^{n-k}-n}{8(n+1)(2n+3)}-1
\]
with clean ancillae and without ancillae, respectively.
\end{theorem}

\begin{proof}
We show that a few branches of a cheap mixture already determine $R$, and that few sets arise this way.  Fix a mixture with clean ancillae, cost $t$ and diamond error $\eps$, with branches $U$ drawn from $\mu$, and put $Q(x)=\E_Up_U(x)$ in the notation of \Cref{lem:acceptance-count}.  On input $\ket x\ket0$, measuring the target gives $1$ with probability $Q(x)$ for the mixture and $\mathbf1_R(x)$ for $U_{\mathbf1_R}$.  The diamond error bounds the trace distance between these outputs, so $\abs{Q(x)-\mathbf1_R(x)}\le\eps$.

Let $\widehat Q(x)$ be the average of $p_{U_i}(x)$ over $N$ independent branches $U_1,\ldots,U_N$.  By Hoeffding's inequality \cite{Hoeffding1963}, for each $x$,
\[
  \Prb\bigl[\abs{\widehat Q(x)-Q(x)}\ge\tfrac12-\eps\bigr]\le2e^{-2N(1/2-\eps)^2}\le2e^{-(n+1)}<2^{-n}.
\]
A union bound over the $2^n$ inputs leaves some sample with $\abs{\widehat Q(x)-\mathbf1_R(x)}<1/2$ for all $x$, and then $R=\{x:\widehat Q(x)\ge1/2\}$.

Hence $R$ belongs to the family $\mathcal F_t$ of sets $\{x:\frac1N\sum_ip_{V_i}(x)\ge1/2\}$, where $V_1,\ldots,V_N$ range over circuits with at most $t$ $T$ gates.  \Cref{lem:acceptance-count} gives $\abs{\mathcal F_t}\le2^{NL}$ with $L=(t+1)(2t+n+3)$.

A random $R$ rarely lands in $\mathcal F_t$.  A uniformly random set of size $2^{n-k}$ is one of $\binom{2^n}{2^{n-k}}\ge2^{k2^{n-k}}\ge2^{2^{n-k}}$ equally likely sets.  Under independent inclusion with probability $2^{-k}\le1/2$, no set has probability above $(1-2^{-k})^{2^n}\le e^{-2^{n-k}}<2^{-2^{n-k}}$.  In both models $\Prb[R\in\mathcal F_t]<2^{NL-2^{n-k}}$.

Take the largest $t$ with $NL<2^{n-k}-n$.  With probability above $1-2^{-n}$, the set $R$ is not in $\mathcal F_t$, so every mixture for it costs more than $t$.  As $NL$ increases with $t$, the first inequality follows.  Without ancillae, the second bound of \Cref{lem:acceptance-count} with $a=0$ replaces $L$ by $(t+1)(2n+3)$ and gives the second inequality.  A union bound over the two cases gives probability $1-2^{1-n}$.  At $\eps=1/4$, the bound $(t+1)(2t+n+3)\le2(t+n+1)^2$ yields the two closed forms.
\end{proof}

The first bound is positive once $n-k\ge3\log n+6$.  The proof uses only that the ancillae start in $\ket0$, so it also covers branches that leave them in other states.

\begin{proposition}[upper bounds for sparse sets]
\label{prop:random-upper}
Let $R\subseteq\F_2^n$ and $\tau=\lceil\log(4\abs R/\eps)\rceil$.  If $\tau\le n$, then with clean ancillae $\Tmix_\eps(U_{\mathbf1_R})=O(\sqrt{\abs R/\eps})$.  If $n\ge\tau+1$, then without ancillae $\Tmix_\eps(U_{\mathbf1_R})\le16\tau\abs R$.
\end{proposition}

\begin{proof}
Hash $x$ by a uniformly random full-rank linear map $H:\F_2^n\to\F_2^\tau$ and accept when $Hx\in H(R)$.  This never errs on $R$.  For $x\notin R$ and $r\in R$, the nonzero vector $x\oplus r$ lies in the kernel of $H$ with probability at most $2^{-\tau}$, so the error at $x$ is at most $\abs R2^{-\tau}$.  As a parity decision tree this gives $\RPDT_\eps(\mathbf1_R)\le\lceil\log(\abs R/\eps)\rceil$ (compare \cite[Fact~B.7]{KMSY18}).

The same hash gives circuits.  A CNOT circuit brings $Hx$ to $\tau$ of the input qubits, the oracle of the $\tau$-bit function $\mathbf1_{H(R)}$ acts there exactly, and the CNOT circuit is undone.  \Cref{lem:pointwise-miss} turns the miss probability into diamond error at most $4\abs R2^{-\tau}\le\eps$.

With clean ancillae the oracle of $\mathbf1_{H(R)}$ costs $O(\sqrt{2^\tau})=O(\sqrt{\abs R/\eps})$ $T$ gates \cite[Theorem~2]{LKS24}, \cite[Lemma~2.1]{GKW24}, as in the randomized sparse QROM of Li et al.\ \cite{LiOuWangYaoYuanZhang2026}.  Without ancillae this oracle is an exclusive-or of at most $\abs R$ point indicators on $\tau$ bits.  Up to $X$ gates each is a gate $C^{\tau}X$, which borrows one of the other $n-\tau\ge1$ inputs and costs at most $16\tau$ $T$ gates (\Cref{sec:mcx}).  This gives $\Tmix_\eps(U_{\mathbf1_R})\le16\tau\abs R$.
\end{proof}

\begin{theorem}[distance does not determine the $T$-count]
\label{cor:distance-not-enough}
Let $n\ge18$ and $2\le k\le n-\lceil2\log(n+1)\rceil-7$.  Let $A$ be an affine subspace of codimension $k$, and let $R$ be random as in \Cref{thm:random-exceptional}.  Then $\delta_H(\mathbf1_A)=2^{-k}$ and $\Tmix_{1/4}(U_{\mathbf1_A})\le24$.  With probability at least $1-2^{2-n}$, also $\delta_H(\mathbf1_R)\in[\frac45,\frac65]2^{-k}$ and $\Tmix_{1/4}(U_{\mathbf1_R})$ is at least
\begin{enumerate}[nosep,label=(\arabic*)]
\item $n-k-\lceil2\log(n+1)\rceil-6$ with any ancillae in any advice state,
\item $\sqrt{(2^{n-k}-n)/(16(n+1))}-n-1$ with clean ancillae, and
\item $(2^{n-k}-n)/(8(n+1)(2n+3))-1$ without ancillae.
\end{enumerate}
If also $k\ge6$, then with the same probability $\Tmix_{1/4}(U_{\mathbf1_R})$ is $O(2^{(n-k)/2})$ with clean ancillae and $O((n-k)2^{n-k})$ without ancillae.
\end{theorem}

\begin{proof}
For $\mathbf1_A$, \Cref{prop:affine-xor} with $J=1$ and $\eps=1/4$ has $\ell_1=\lceil\log16\rceil=4$.  Since $n\ge5$, it gives $c(\min\{k,4\})\le c(4)=24$ $T$ gates.  The distance $\delta_H(\mathbf1_A)=2^{-k}$ holds because $k\ge2$.

For $\mathbf1_R$, we apply \Cref{thm:random-exceptional} with $s=\lceil2\log(n+1)\rceil+6$, which satisfies \eqref{eq:s-condition}.  The upper bound on $k$ gives $q=n-k-s\ge1$, so the theorem bounds $\delta_H(\mathbf1_R)$ and gives $\Tmix_{1/4}(U_{\mathbf1_R})\ge n-k-s=n-k-\lceil2\log(n+1)\rceil-6\ge1$.  The bounds with clean ancillae and without ancillae are \Cref{thm:random-exponential}, and a union bound over the two theorems gives the probability.

For the upper bounds, the same event gives $\abs R\le\frac65 2^{n-k}$, so at $\eps=1/4$ we have $\tau=\lceil\log(16\abs R)\rceil\le n-k+5$.  When $k\ge6$ this is at most $n-1$, and \Cref{prop:random-upper} gives the two bounds.
\end{proof}

Together, the $T$-count of a random set is $2^{(n-k)/2}$ with clean ancillae and $2^{n-k}$ without, both up to factors polynomial in $n$.  Decision trees cannot show this, since $\RPDT_{1/4}(\mathbf1_R)\le\lceil\log(4\abs R)\rceil\le n-k+3$.

\section*{Acknowledgments}

The author thanks Ryan O'Donnell, who introduced the author to this topic, engaged in helpful discussions, and suggested the one-clean-ancilla construction in part 3 of \Cref{prop:toffoli-borrowed}.  The author also thanks Junseo Lee for detailed feedback on earlier drafts.
\section*{AI Disclosure}

Generative artificial intelligence tools (Claude) assisted the author in deriving and refining some of the proofs, in writing the verification code, and in drafting and revising the manuscript, including its structure, cross-references and comparisons with cited sources.  The author reviewed and verified all AI-assisted arguments and text and takes full responsibility for the content of this work.

\bibliographystyle{unsrt}
\bibliography{shared/references1}

\end{document}